\documentclass[]{jfm}

\usepackage{graphicx}
\usepackage{array}
\usepackage{newtxtext}
\usepackage{newtxmath}
\usepackage{natbib}
\usepackage[hoptionsi]{overpic}
\usepackage[table,xcdraw,dvipsnames]{xcolor}
\usepackage{amsmath}
\usepackage{mathtools}
\usepackage{hyperref}
\hypersetup{
    colorlinks = true,
    urlcolor   = blue,
    citecolor  = blue,
}
\usepackage[T1]{fontenc}
\usepackage{amsfonts}
\usepackage{amsmath}
 \usepackage{mathtools}
\usepackage{cuted}
\usepackage{cuted} 
\usepackage{relsize}

\newcommand{\RomanNumeralCaps}[1]
\linenumbers
\usepackage{soul}

\title{A fast-running physics-based wake model for a semi-infinite wind farm}

\author{Majid Bastankhah\aff{1}
  \corresp{\email{majid.bastankhah@durham.ac.uk}},
 Mohammad Mehdi Mohammadi\aff{2},
 Charlie Lees\aff{1},
 Gonzalo Pablo Navarro Diaz\aff{2},
 Oliver Buxton \aff{3}
 \and Stefan Ivanell \aff{2}}

\affiliation{\aff{1}Department of Engineering, Durham University, Durham DH1 3LE, UK
\aff{2}Department of Earth Sciences, Uppsala University, Uppsala, Sweden
\aff{3} Department of Aeronautics, Imperial College London, SW7 2AZ, UK}

\begin{document}
\maketitle

\begin{abstract}
This paper presents a new generation of fast-running physics-based models to predict the wake of a semi-infinite wind farm, extending infinitely in the lateral direction but with finite size in the streamwise direction. The assumption of a semi-infinite wind farm enables concurrent solving of the laterally-averaged momentum equations in both streamwise and spanwise directions. The developed model captures important physical phenomena such as vertical top-down transport of energy into the farm, variable wake recovery rate due to the farm-generated turbulence, and also wake deflection due to turbine yaw misalignment and Coriolis force. Of special note is the model's capability to predict and shed light on the counteracting effect of Coriolis force causing wake deflections in both positive and negative directions. Moreover, the impact of wind-farm layout configuration on the flow distribution is modelled through a parameter called the local deficit coefficient. Model predictions were validated against large-eddy simulations extending up to 45 kilometres downstream of wind farms. Detailed analyses were performed to study the impacts of various factors such as incoming turbulence, wind-farm size, inter-turbine spacing, and wind-farm layout on the farm wake.
\end{abstract}

\section{Introduction}

Offshore wind is projected to experience rapid expansion in the coming decades, emerging as a significant global renewable energy source \citep{veers2019}. To achieve this goal, many new offshore wind farms are anticipated to be erected in specific and promising geographical areas, particularly in regions like the North Sea, where strong and consistent winds are present. Consequently, the interaction among neighboring offshore wind farms, as their wakes affect each other, has become an essential and pressing subject of research. Recent satellite images and field measurements have revealed that wakes of wind farms can last for many kilometres \citep{christiansen2005,nygaard2018wake,ahsbahs2020wind}. Significant power degradation and fatigue loads can thus occur for a wind farm subject to wakes of adjacent wind farms \citep{stevens17}. Beyond technical complexities, interactions between adjacent wind farms may lead to legal and financial disputes between operators of neighboring facilities. As a result, accurate and reliable modelling of wind-farm wake effects becomes of great importance for optimizing future wind farms in increasingly competitive offshore environments.


High-fidelity numerical simulations such as large-eddy simulation (LES) are powerful tools for modeling complex turbulent wake flows, offering detailed insights into flow dynamics and wake interactions \citep[][and references therein]{porte2020}. However, simulating a cluster of wind farms in congested areas such as the North Sea with LES is computationally intensive and time-consuming, making it impractical for real-time or large-scale studies. To address this challenge and enable more efficient simulations, there is a clear demand for fast-running engineering wake models striking a balance between accuracy and computational cost.  Major advantages of these models are their ease of use and low computational costs, allowing for quicker assessments of various scenarios and aiding in optimization of wind farm layouts and real-time control. Below, we attempt to classify the engineering wake models developed in the literature.


The typical method for modeling airflow distribution within wind farms involves predicting the wake generated by each individual turbine. A superposition method is then applied to consider the combined impact of these wake effects. These individual wake models range mainly from top-hat models \citep{Jensen1983,Katic1986} to Gaussian-type models \citep{bastankhah14}. The Jensen top-hat model (also known as the Park model) has been extended in recent works to account for variable wake recovery rate due to turbine-generated turbulence \citep{nygaard2020modelling}. Over time, Gaussian wake models have also been refined and extended in several studies to more accurately describe the near-wake region \citep[e.g.,][]{keane2016,shapiro2019paradigm,blondel2020,schreiber2020brief}, to better capture wake expansion and its asymmetrical shape \citep[e.g.,][]{xie2014self,abkar2015influence,vahidi2022physics,pedersen2022turbulence} or to capture effects of yaw angle \citep[e.g.,][]{bastankhah2016experimental,king2020,bastankhah2022vortex, bay2023} and wind veer \citep{abkar2018analytical,mohammadi2022veer,narasimhan2022}. Moreover, a variety of wake superposition methods exist, aiming to model cumulative wake effects in wind farms \citep[e.g.,][]{Lissaman1979, voutsinas1990, Niayifar2016, zong2020, bastankhah2021analytical,lanzilao2022superposition}. Some of these methods are solely empirical in nature, while others have a
foundation in flow physics. See \cite{bastankhah2021analytical} for a detailed discussion on different wake superposition methods. This simple approach has proven to be very useful in providing detailed information on the flow field within small-sized wind farms and has been extensively used in wind farm layout optimisation and real-time flow control; see the review of \cite{meyers2022review} and references therein. However, this modelling approach cannot properly describe the interaction of wind farms with the atmospheric boundary layer (ABL) which involves scales that are comparable to the size of the entire wind farm or the ABL thickness. Most notably, these models fall short of capturing the crucial vertical transport of kinetic energy from higher altitude layers of the atmosphere into the wind farm/wind-farm wake \citep{stevens17}. This becomes especially  problematic as the size of wind farms grows, or if we seek information about the wake of the entire wind farm several kilometres downstream.

Capturing large-scale wind farm physics may be more readily achieved using \emph{infinite wind farm models}. In this approach, the wind farm is assumed to be infinitely large in both lateral and streamwise directions, and the whole wind farm is modelled as an area with an increased aerodynamic surface roughness. Unlike single-wake modelling, this approach is able to capture the vertical transport of energy caused by turbulent fluxes, which is in balance with the energy extracted by wind turbines in infinite wind farms. The interested reader is referred to the seminal works of  \cite{frandsen1992}, \cite{calaf2011} and other subsequent studies \citep[][amongst others]{Frandsen2006,yang2012,meneveau2012,meyers-meneveau2012,abkar2013,stevens2016effects} for more information.
Despite the great advantage of these models in capturing the farm-atmosphere interaction, the concept of an infinite wind farm can be only regarded as an asymptotic case that resembles what very large wind farms may tend to approach. More importantly, these models fail to offer any insight into the wake of the wind farm due to their core assumption that the wind farm extends infinitely in the streamwise direction.

The other group of existing models, which we classify within the broad category of the \textit{multi-scale} models, strive to leverage the benefits of both large-scale farm and small-scale single turbine modelling. Within this category, different approaches have been adopted to model wind farm flows. \cite{stevens2015} and subsequent works \citep[e.g.,][]{shapiro2019paradigm,starke2021} coupled the infinite-farm approach with the single-turbine approach by matching the predicted mean velocity at the turbine hub height. Other studies coupled the wind-farm scale with the turbine-scale through a parameter called the farm induction factor \citep[e.g.,][]{nishino2020two,kirby2022two} with more recent works modelling blockage effects as well \citep[e.g.,][]{legris2022}. In another type of multi-scale model, the exchange of energy between the layer consisting of wind turbines and the overlaying boundary layer was parameterised using the classical entertainment theory \citep[e.g.,][]{luzzatto2018,bempedelis2023}. Other multi-scale models characterised the farm-atmosphere interaction and farm-scale blockage effects caused by meso-scale phenomena such as gravity waves \citep{allaerts2019sensitivity, stipa2023Allaert}. The coupling between the different scales in these models usually involves an iterative process or the numerical solution of governing equations. Moreover, the focus of the majority of the models discussed above is farm power production or the flow field within the wind farm, and less attention has been paid to the wake of the entire farm.

\begin{figure}
    \centering
    \includegraphics[width=\textwidth]{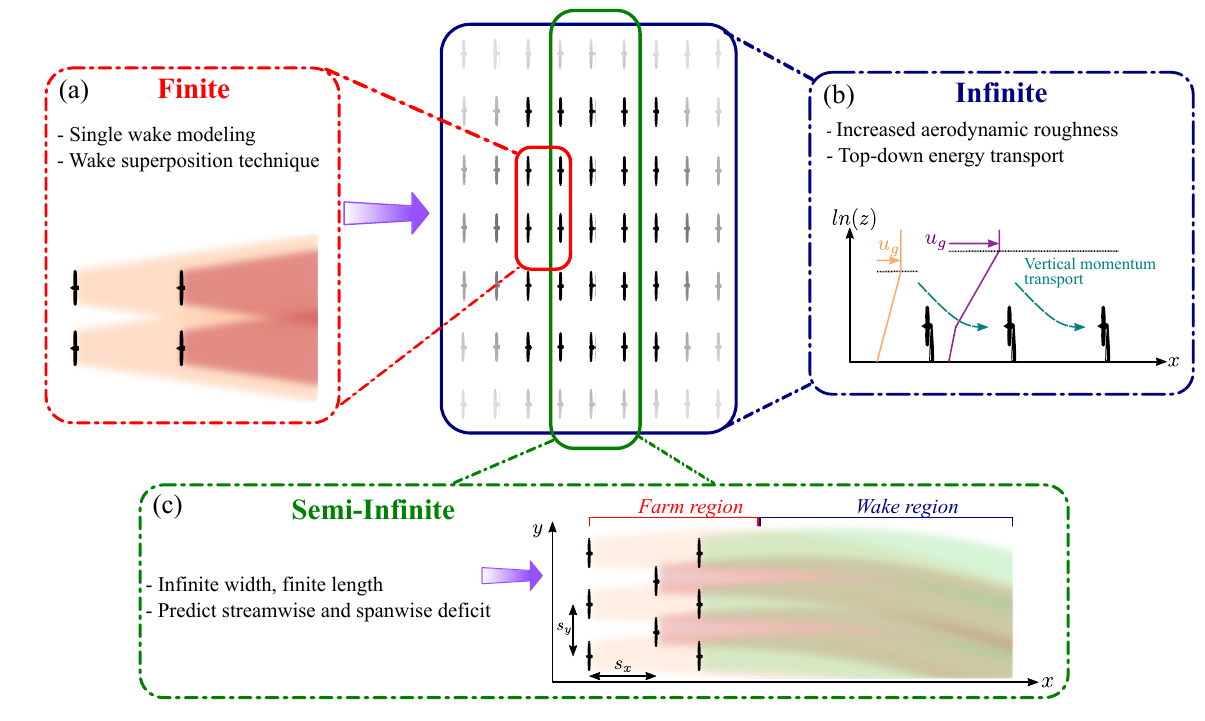}
    \caption{Different approaches used to model wind farm flows: (a) modelling each individual wind turbine wake and then using superposition techniques to account for cumulative wake effects, (b) modelling wind farm that is extended to infinity in both streamwise and spanwise directions as an added aerodynamic surface roughness, (c) modelling a wind farm that is extended to infinity in the lateral direction but has a finite length in the streamwise direction. }
    \label{fig:schematic_semi-infinite}
\end{figure}

In this work, we propose a new category of wake models by considering a \textit{semi-infinite} wind farm; i.e., a wind farm that extends infinitely in the lateral direction but has a finite size in the streamwise direction. The infinite lateral extent of the wind farm allows us to perform lateral averaging, which significantly simplifies the flow's governing equations and leads to a closed-form explicit solution without the need for using an iterative approach. The finite length of the wind farm also makes it possible to systematically model the wake of the entire farm. A schematic of the semi-infinite farm modelling in comparison with the single-turbine modelling (i.e., finite approach) as well as the infinite-farm modelling is shown in figure \ref{fig:schematic_semi-infinite}.  A particular focus of this work is given to the prediction of the deflection of the farm wake. Predicting the magnitude of the farm wake deficit is important but not sufficient. The wake deflection also needs to be quantified to determine whether the wake of a wind farm may impinge on a downstream farm. In general, the wake deflection is mainly caused by (i) meso-scale phenomena such as Coriolis force (and its by-product wind veer), and (ii) yaw misalignment. The latter has recently received a great deal of attention because of  its importance in wake steering strategies \cite[][amongst others]{fleming2017field,howland2019,bastankhah2019wind,campagnolo2020wind}. The former, however, is mostly overlooked in prior modelling works. While the deflection of a single turbine wake due to Coriolis force is expected to be negligible \citep{mohammadi2022veer}, several studies mainly based on numerical simulations have underpinned the importance of the farm-wake deflection caused by Coriolis force \citep[e.g.,][]{van2015predicting,abkar2016wake, allaerts2017,gadde2019Stevens,eriksson2019ivanell}. Interestingly, there has not been a universal agreement in the literature with regards to the direction of the wake deflection caused by the Coriolis force. \cite{vanderLaan2017WhyHemisphere} argued that this is due to the fact that the Coriolis force has two effects on the wake deflection. The direct effect turns the wake in the anticlockwise direction (seen from top) in the Northern hemisphere, while the indirect effect (through wind veer) rotates the wake in the clockwise direction. \citet{gadde2019Stevens} explained this phenomenon based on the direction of the vertical turbulent fluxes in the entrance region in comparison with those in the wake. One of our objectives with this new modelling framework is to capture the conflicting influence of the Coriolis force on the farm wake. This is achieved by concurrently solving momentum equations in both streamwise and spanwise directions. The outcome is a simple one-dimensional model that predicts both laterally-averaged streamwise and spanwise velocities at the turbine hub height within and downwind of the wind farm. 


The rest of the paper is organised as follows. Section \ref{sec:LA-RA-NSE} develops the laterally-averaged Reynolds-averaged Navier-Stokes equations. Section \ref{sec:LES_setup} describes the high-fidelity numerical simulations used in this study to validate the developed model. Section \ref{sec:budget_analysis} discusses the budget analysis that is conducted to identify dominant terms in the momentum equations. The farm-wake model is then developed in section \ref{sec:model_development}. Results are discussed in section \ref{sec:results}, and finally a summary is provided in section \ref{sec:summary}.

\section{Streamwise and Spanwise Laterally-Averaged Reynolds-Averaged Navier-Stokes Equations} \label{sec:LA-RA-NSE}

We start by writing the steady-state Reynolds-averaged Navier-Stokes (i.e., RANS) equation for high-Reynolds flows (i.e., negligible friction forces) using Einstein notation. For simplicity, we non-dimensionalise all variables and equations throughout this manuscript using a selection of scales based on the incoming flow and turbine characteristics. All spatial dimensions are normalised by the turbine rotor diameter $D$. All velocities $u_i$ are normalised by the incoming velocity $\mathcal{U}_h$ at the turbine hub height $z_h$. Static pressure $p$ is normalised by $\rho \mathcal{U}_h^2$, where $\rho$ is the air density. The dimensionless Coriolis frequency $f_c$ is defined as
\begin{equation}\label{eq:f_c}
    \frac{Df_c}{\mathcal{U}_h}=2\Omega\sin\phi,
\end{equation}
 where $\Omega= 7.2921 × 10^{-5} $rad/s is the rotation rate of the earth, and $\phi$ is the latitude. Note that the dimensionless Coriolis frequency $f_c$ defined in \eqref{eq:f_c} represents the ratio of Coriolis force to inertial force. This dimensionless parameter is in fact the inverse of the Rossby number $Ro$ that is commonly used in geophysical studies concerning flows in oceans and atmosphere \citep[e.g.,][]{van2020rossby}.  
 
 The dimensionless form of the RANS equation reads as   \citep{stull2009}:  
\begin{align}\label{eq:NSx}
\bar{u}_j\frac{\partial \bar{u}_i}{\partial x_j}  &=\varepsilon_{ij3}f_c\bar{u}_j-\frac{\partial \bar{p}}{\partial x_i}- \frac{\partial \overline{u_i'u_j'}}{\partial x_j}-\bar{f}_i,
\end{align}
where $u_i$ is the velocity component in the $x_i$ direction with $i = 1, 2, 3$ corresponding to the streamwise $x$, spanwise $y$ and vertical $z$ directions, respectively. Overbar denotes time averaging, and the turbulent fluctuating velocity is $u_i'=u_i-\bar{u}_i$. The permutation symbol is denoted by $\varepsilon_{ijk}$. 
Moreover, $\bar{f}_i$ is the time-averaged component of the turbine forces per unit volume, non-dimensionalised by $\rho \mathcal{U}_h^2/D$.  Note that gravitational forces are  neglected in \eqref{eq:NSx} as only the streamwise and spanwise momentum directions are of interest in this study.

We separate the static pressure $p$ in \eqref{eq:NSx} into the one due the background driving pressure gradient $p_g$ and the one due to the presence of turbines $p_t$. The former is dictated by the force balance in the geostrophic layer on top of the ABL, while the latter is due to the pressure drop across the rotor disk and its recovery to the free-stream pressure downstream. The background driving pressure gradient $p_g$ can be written in terms of the geostrophic wind speed $\bar{u}_g$ \citep{stull2009}. This simplifies \eqref{eq:NSx} to
\begin{align}\label{eq:NS2}
\bar{u}_j\frac{\partial \bar{u}_i}{\partial x_j}  &=-\varepsilon_{ij3}f_c(\bar{u}_{gj}-\bar{u}_j)-\frac{\partial \bar{p}_t}{\partial x_i}- \frac{\partial \overline{u_i'u_j'}}{\partial x_j}-\bar{f}_i.
\end{align}

Different forms of spatial averaging such as surface or volumetric averaging have been frequently performed in prior studies on vegetation canopies or infinite wind farms \citep[e.g.,][]{calaf2010large,Moltchanov2011DispersiveEdge, bai2015meneveau,Goit2015OptimalLayers}. In this study, however, we perform lateral averaging (averaging only along the $y$ direction), because our aim is to determine how flow quantities at the hub-height level evolve along the streamwise direction $x$. Mathematically, this may be defined for an arbitrary variable $\psi$ as follows
\begin{equation}\label{eq: lateral avg}
    \langle  \bar{\psi} \rangle (x,z)= \lim_{L\to\infty}\frac{1}{2L}\int_{-L}^L\bar{\psi}(x,y,z) dy,
\end{equation}
where $\langle\rangle$ indicates lateral averaging, and $[-L,L]$ is the lateral range over which the averaging is performed.  The lateral fluctuation is defined as $\psi''=\psi-\langle \psi \rangle$ and by definition $\langle \psi''\rangle=0$. By performing the lateral average on \eqref{eq:NS2}, we obtain

\begin{align}\label{eqn:LANS}
\begin{aligned}
    \underbrace{\langle \overline{u_{j}}\rangle \frac{\partial \langle \overline{u_{i}}\rangle}{\partial x_{j}}}_{\mathlarger{A}} = 
    -&\underbrace{\varepsilon_{\mathit{ij3}}f_c (\langle \overline{u_{\mathit{gj}}}\rangle  -\langle \overline{u_{j}}\rangle)\vphantom{\frac{\rangle\partial}{\partial}}}_{\mathlarger{C}} - \underbrace{{\frac{\partial \langle\overline{p_t}\rangle}{\partial x_i}}}_{\mathlarger{P}} 
     - \underbrace{\frac{\partial \langle \overline{u_{i}' u_{j}'}\rangle}{\partial x_{j}}}_{\mathlarger{R}}
    - \underbrace{\frac{\partial \langle \overline{u_{i}}'' \overline{u_{j}}'' \rangle}{\partial x_{j}}}_{\mathlarger{D}}-\underbrace{\langle  \bar{f}_i \rangle}_{\mathlarger{T}}.
    \end{aligned}
\end{align}
The terms outlined in \eqref{eqn:LANS} are 
\begin{itemize}
    \item [A] Advection of momentum by mean flow 
    \item [C] Coriolis term
    \item [P] Pressure gradient due to the presence of wind turbines
    \item [R] Reynolds stress gradients
    \item [D] Dispersive stress gradients
    \item [T] Turbine forcing
\end{itemize}
These terms will be discussed in more details in section \ref{sec:budget_analysis}. The dispersive stress term $\langle \overline{u_{i}}'' \overline{u_{j}}'' \rangle$ that is the product of spatial fluctuations in the lateral direction arises in \eqref{eqn:LANS} as a result of lateral averaging. It is also noteworthy that because of lateral averaging any terms including $\partial \langle \rangle/\partial x_j$ in \eqref{eqn:LANS} must be zero if $j=2$ (i.e., $\partial/\partial y=0$).  
 
\section{Numerical setup: Large-eddy simulation (LES)}\label{sec:LES_setup}



LESs were performed using the open source software OpenFOAM (version 2.3.1) in conjunction with the Simulator fOr Wind Farm Applications (SOWFA) project libraries \citep{churchfield_large-eddy_2012} developed by the U.S. National Renewable Energy Laboratory (NREL). 
The atmospheric solver used in SOWFA is called ABLSolver, 
which is a transient solver for turbulent flows of incompressible fluids and considers the Boussinesq approximation for buoyancy effects \citep{churchfield_large-eddy_2012}. 
%

A precursor-successor approach has been utilized to develop a conventionally neutral atmospheric boundary layer flow for the simulation. The Coriolis force is calculated for $\phi=55.52^{\circ}$, which is a representative value for a wind farm in the North Sea \citep{hansen2012}. 
In SOWFA, a prescribed streamwise velocity ($\mathcal{U}_h=8$ m/s) and wind direction ($\varphi=270^{\circ}$) at the turbine hub-height level can be achieved by adjusting the magnitude and direction of the driving pressure gradient.
A capping inversion with a lapse rate of 0.05 K/m is imposed at the top of the boundary layer covering the heights from  $700$ m to $800$ m. The height at the bottom of the capping inversion, denoted by $H$, is defined as the thickness of the ABL. The geostrophic layer above the capping inversion has a lapse rate of 0.003 K/m. The inclusion of the capping inversion helps to slow the vertical growth of the boundary layer with time in neutral conditions \citep{churchfield2012numerical}. Due to the assumption of the fixed height of the capping inversion, however, the vertical displacement of the flow above the farm does not generate gravity waves in the capping inversion. This simplification may lead to errors in cases where gravity waves induce non-negligible pressure gradients at the turbine hub-height level  \citep{allaerts2017,allaerts2019sensitivity,stipa2023Allaert,lanzilao2023}.
The precursor simulations are run without the turbines for a period of 10 hours (36,000 seconds) to obtain a quasi-steady state. Next, the inlet conditions are recorded for a period of 9,000 seconds to be fed into the successor simulation with turbines. The farm flow statistics are calculated for the last hour of the simulations. The convective terms are discretised using a second order central difference scheme for the precursor and a local blend between linear (second order) and upwind (first-order) schemes for the successor simulation depending on the cell size. This scheme uses 80\% linear and 20\% upwind in proximity of the turbines and 100\% linear in the rest of the domain. For temporal discretisation, an implicit second order backward scheme is used. For the diffusion term, a Gauss linear second-order scheme is implemented using a non-orthogonality correction for surface normal gradients. Subgrid-scale (SGS) stresses are modelled using a one-equation turbulent-viscosity model
%
\citep{yoshizawa1986statistical}.

The turbine used in this study is NREL-5MW with a hub height of 90 m and a rotor diameter $D$ of 126 m \citep{osti_947422}. The turbines are modeled as an actuator disk with no rotation and a constant thrust coefficient of 0.776. This value of $C_T$ was found based on blade element momentum (BEM) simulations of a turbine rotor with $\mathcal{U}_h$ of 8 m/s and the tip-speed ratio of 7.55 \citep{navarro2022actuator}. The body forces are spread across the rotor plane uniformly as axial forces. The equivalent inflow velocity is unknown for turbines that are subject to the wakes of upstream turbines, so a calibration table is used to relate the average velocity on the disc with the unperturbed inflow velocity \citep{Laan2015}.

\begin{figure}
\centering
\includegraphics[width=1\textwidth]{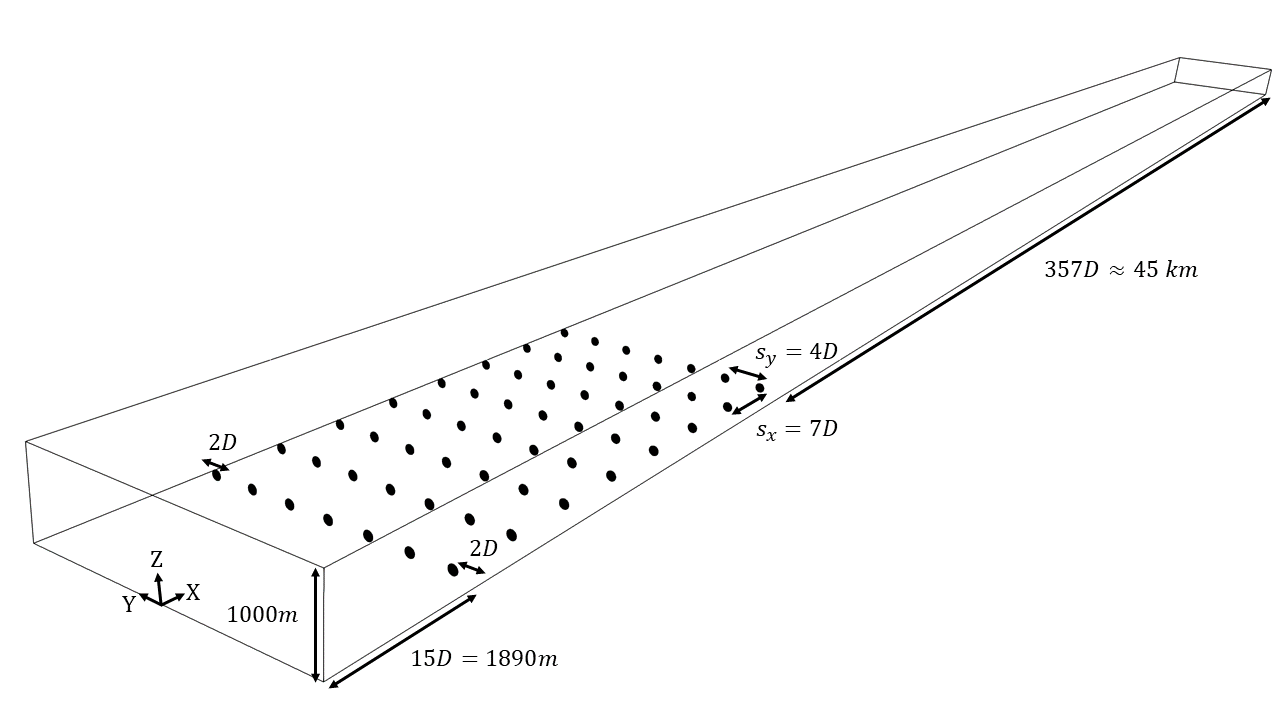}
\caption{\label{fig:domainpic}Schematic of the computational domain for the A0 case. Turbines are shown by black circles, and the rotor diameter is denoted by $D$. }
\end{figure}

The same domain size is used for both precursor and successor simulations. It extends 1,000 m in the vertical direction. The first row of turbines is placed 15 rotor diameters (1,890 m) downstream of the inlet, and the domain extends for 357 turbine diameters (approximately 45 km) after the last turbine row. A schematic of the computational domain is presented in figure \ref{fig:domainpic}.  It is worth noting that the distance between the inlet and the first row of turbines is relatively short in these simulations. Our aim is to maximize the use of our computational resources to capture a very large extent of the farm wake. However, this may lead to an underestimation of the velocity slowdown in the upwind region caused by farm-scale blockage effects, as discussed in the recent parametric study by \cite{lanzilao2023}.

%
In the precursor simulations, grid cells are 21 m  (i.e., $D/6$) long in the streamwise and lateral directions, and their height in the vertical direction grows with distance from the ground (from 2.5 m to 60 m at the top of the domain). In the successor simulations including wind turbines, the mesh is refined in two steps. Each refinement halves the cell size. First, in a zone containing the wind farm and its downstream region, the mesh size is reduced to 10.5 m (i.e., $D/12$) in the streamwise and lateral directions.  This refined region starts 1,260 m (i.e., $10D$) upstream of the first turbine row to ensure that eddy structures are fully developed in the new refined mesh before reaching the wind turbines. In close proximity to the wind turbines, the mesh is further refined by a factor of two (i.e., cell size of 5.25 m or  $D/24$ in the streamwise and lateral directions) to capture strong velocity gradients in this region.

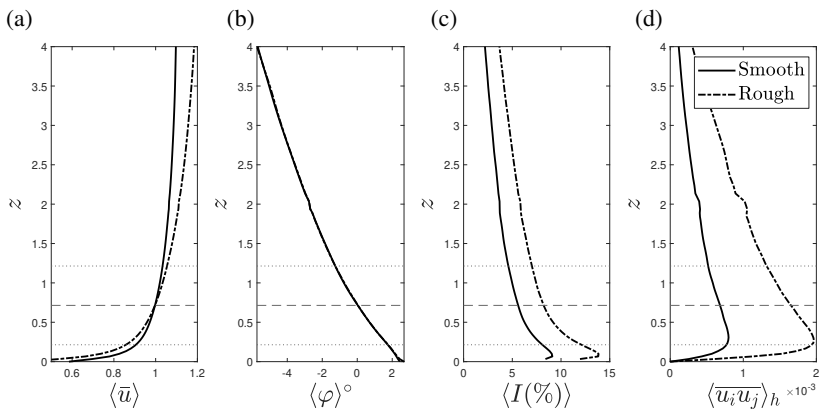
\begin{figure}
    \centering
       \begin{overpic}[width=.8\textwidth]{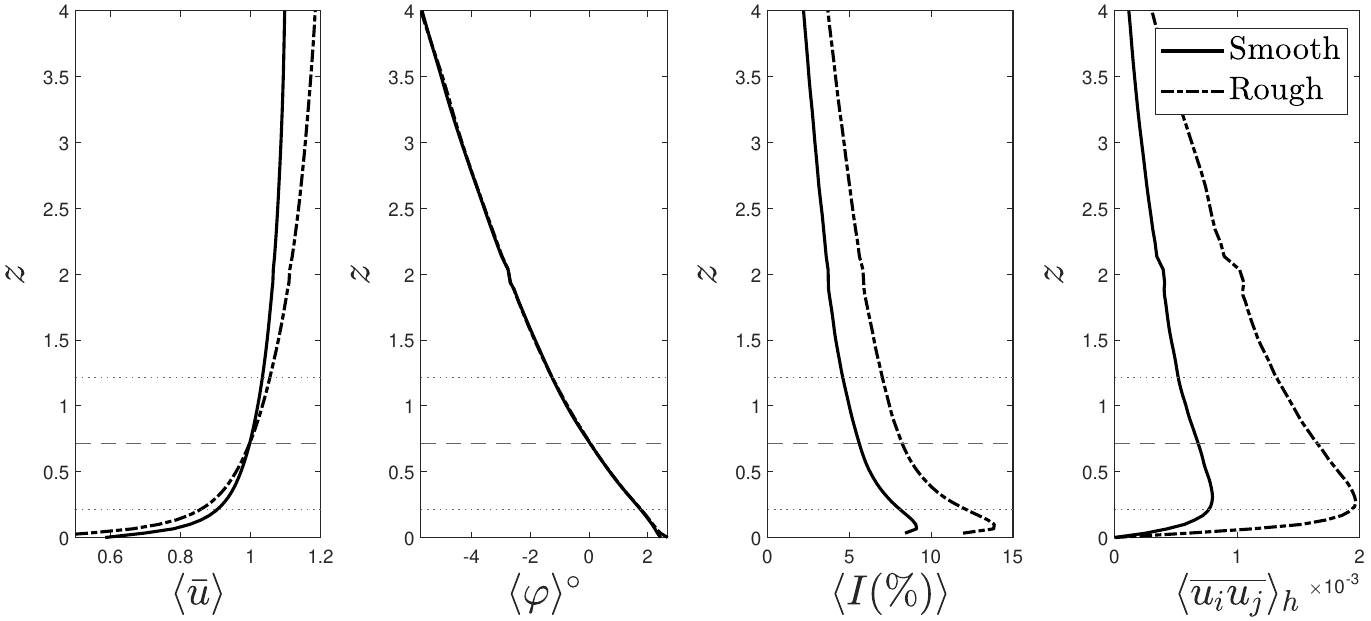}
    \put(0,47){(a)}
    \put(27,47){(b)}
      \put(52,47){(c)}
    \put(77,47){(d)}
    \end{overpic}
       \caption{Spanwise-averaged vertical profiles of inflow characteristics obtained from precursor simulations: (a) the normalised streamwise velocity $\langle\bar{u}\rangle$ (b) the wind direction $\langle\varphi\rangle$, (c) the incoming turbulence intensity $I = \sigma_u/\mathcal{U}_h$, where $\sigma_u$ is the standard deviation of streamwise turbulent fluctuations, and (d) the horizontal turbulent shear stress defined as $\langle\overline{u_iu_j}\rangle_h=\sqrt{\langle \overline{u'w'}\rangle^2+\langle \overline{u'v'}\rangle^2}$. Horizontal dashed and dotted lines respectively indicate the turbine hub height and vertical positions of top/bottom blade tips. }
    \label{fig:inflow}
\end{figure}
\begin{figure}
\centering
\includegraphics[width=1\textwidth]{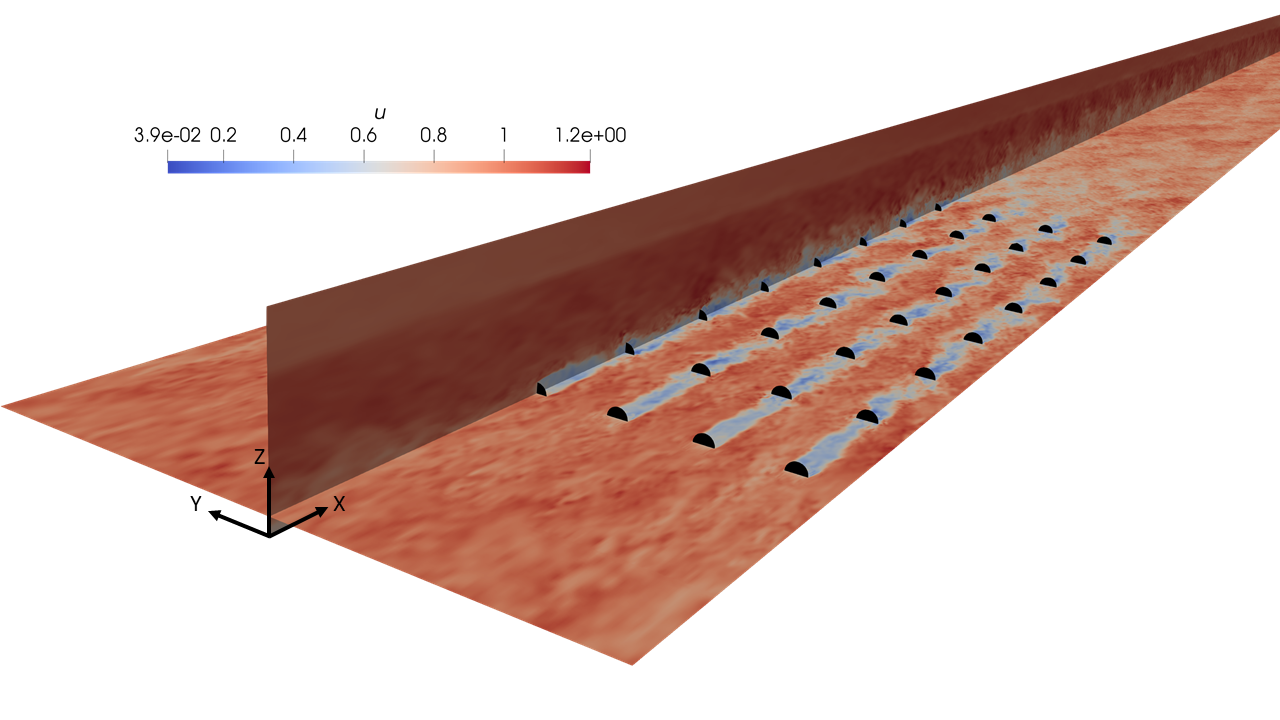}
\caption{\label{fig:3dview}Contours of instantaneous normalised streamwise velocity $u$ for the Aligned Baseline (A0) case. Turbines are shown by black circles. }
\end{figure}

Precursor simulations use cyclic boundary conditions at the inlet, outlet, and sides. The nearest turbines to the sides are placed such that it resembles the infinite extent of the wind farm in the lateral direction. For instance, with a lateral spacing of 4D, there is a 2D distance from each side as shown in figure \ref{fig:domainpic}. At the ground, a wall boundary condition with a prescribed roughness length based on the Schumann Grotzbach formulation  is implemented \citep{schumann1975subgrid}.  At the domain top, a slip boundary condition is imposed for velocity and a fixed gradient for temperature. For successor simulations including wind turbines, the inlet uses the data from the precursor while a zero-gradient condition is applied at the outlet. The domain's sides, lower, and upper parts have cyclic, wall, and slip boundary conditions, respectively.  
In total, five simulations were performed to study the effect of wind farm layout, wind farm length, inter-turbine spacing and the incoming turbulence level on farm wake flows. The details of these simulations are summarised in Table \ref{tab:runs}. In this table, $s_x$ and $s_y$ are, respectively, streamwise and spanwise inter-turbine spacing, normalised by the rotor diameter $D$. The surface roughness normalised by $D$ is shown by $z_{0,0}$. The number of turbine rows is denoted by $N$, and $u_*$ is the friction velocity normalised by $\mathcal{U}_h$. The incoming turbulence intensity at the hub height is shown by $I_0$, and $\Delta \varphi$ is the change in the incoming wind direction across the turbine rotor (i.e., from the bottom-tip height to the top-tip height). As seen in Table \ref{tab:runs}, the first four cases (A0, S0, AS, AD) are subject to a smooth boundary layer with low surface roughness, whereas the incoming boundary layer in the AR case has a higher surface roughness. The two different inflow boundary-layer profiles are shown in figure \ref{fig:inflow}. The instantaneous streamwise velocity field $u$ for a portion of Aligned Baseline (A0) case is also shown in figure \ref{fig:3dview}, where the highly-turbulent nature of the atmospheric flow and low-speed wakes are clearly visible.

\begin{table}
\begin{center}
\begin{tabular}{m{3 em} m{8em} m{5 em}m{1 em}m{1 em}m{5 em}m{1 em} m{4 em} m{3 em} m{3 em}} 
\hline\hline
case & description        & layout    & $s_x$ & $s_y$ & $z_{0,0}$               & $N$ & $u_*$ & $I_0(\%)$ & $\Delta \varphi(^\circ)$ \\ \hline
A0   & Aligned Baseline   & aligned   & 7     & 4     & $1.6\times 10^{-6}$ & 8   & 0.0037                       &      5.6    &  -3\\
S0   & Staggered Baseline & staggered & 7     & 4     & $1.6\times 10^{-6}$ & 8   &  0.0037                         &       5.6  &  -3\\
AS   & Aligned Short      & aligned   & 7     & 4     & $1.6\times 10^{-6}$ & 4   &  0.0037                         &       5.6  &  -3\\
AD   & Aligned Dense      & aligned   & 5     & 3     & $1.6\times 10^{-6}$ & 8   &  0.0037                         &        5.6  & -3\\
AR   & Aligned Rough      & aligned   & 7     & 4     & $1.6\times 10^{-4}$ & 8   &   0.0059                        &        8.2  & -3\\ \hline
\end{tabular}
\end{center}
\caption{Summary of LES for different semi-infinite wind farms. $s_x$ and $s_y$ are, respectively, streamwise and spanwise inter-turbine spacing, normalised by the rotor diameter $D$. The surface roughness normalised by $D$ is shown by $z_{0,0}$. The number of turbine rows is denoted by $N$, and $u_*$ is the friction velocity normalised by $\mathcal{U}_h$. The incoming turbulence intensity at the hub height is shown by $I_0$, and $\Delta \varphi$ is the change in the incoming wind direction across the turbine rotor (i.e., from the bottom-tip height to the top-tip height). 
}\label{tab:runs}
\end{table}

\section{Momentum Budget Analysis}\label{sec:budget_analysis}
In this section, the LES data for the Aligned Baseline (A0) case are employed to perform budget analysis on the laterally-averaged momentum equations \eqref{eqn:LANS}. This analysis determines dominant terms in the momentum equations both within and downwind of the wind farm. This serves as a basis for the development of the physics-based model later in section \ref{sec:model_development}. Note that viscous terms in the momentum equations were found to be multiple orders of magnitude smaller than any other term, so they are neglected in this analysis. Moreover, regions immediately downstream and upstream of the turbine rows are removed due to steep flow gradients in these regions. This analysis investigates all terms in \eqref{eqn:LANS}, except for the turbine forcing which is only relevant at the rotor disk.
\subsection{Streamwise momentum equation}\label{sec:budget_analysis_streamwise}



\begin{figure}
    \centering
    \includegraphics[width =.9\textwidth]{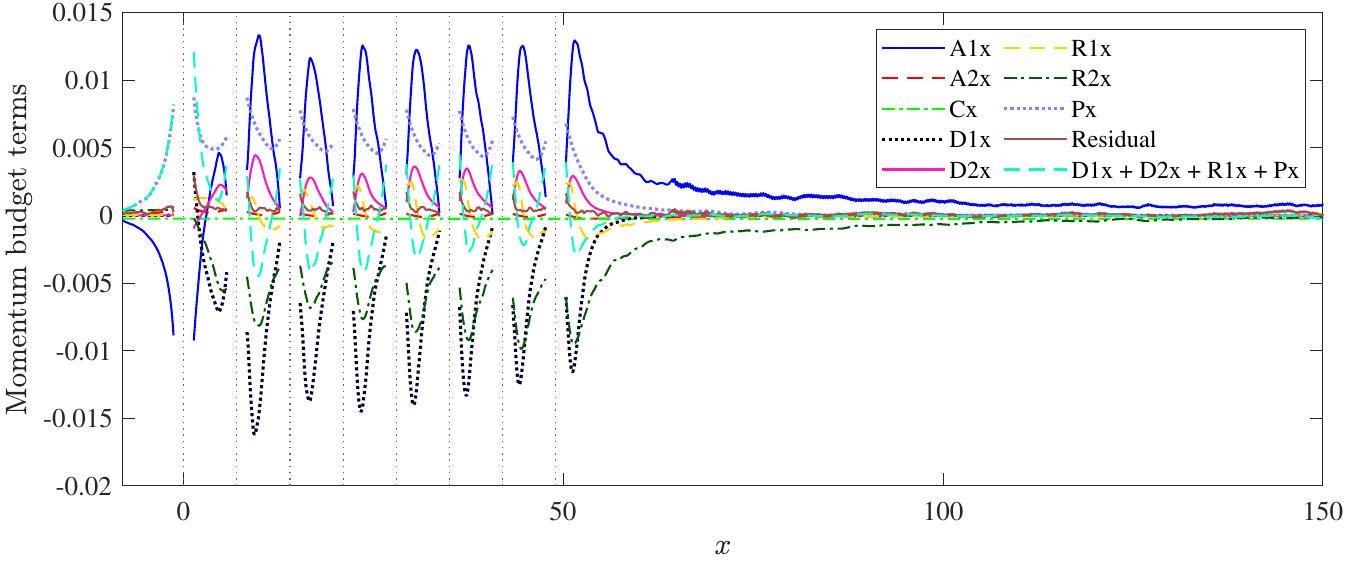}
    \caption{Non-dimensionalised streamwise momentum \eqref{eqn:DANSx} budget at hub height for the Aligned Baseline (A0) case. Locations of turbine rows are denoted by vertical dotted lines. All variables are non-dimensionalised 
using a selection of $\mathcal{U}_h$ and $D$.}
    \label{fig:budget-streamwise}
    \vspace{-0.5cm}
\end{figure}




Writing \eqref{eqn:LANS} in the streamwise direction (i.e., $i=1$) and neglecting turbine forcing gives 
\begin{align} \label{eqn:DANSx}
\begin{aligned}
    \underbrace{\langle\overline{u}\rangle\frac{\partial\langle\overline{u}\rangle}{\partial x}}_{\mathlarger{A1x}} + \underbrace{\langle\overline{w}\rangle \frac{\partial\langle\overline{u}\rangle}{\partial z}}_{\mathlarger{A2x}} = -\underbrace{f_c(\langle\overline{v_g}\rangle - \langle\overline{v}\rangle)\vphantom{\frac{\partial}{\partial}}}_{\mathlarger{Cx}} - \underbrace{\frac{\partial\langle\overline{u}''\overline{u}''\rangle}{\partial x}}_{\mathlarger{D1x}}
    - \underbrace{\frac{\partial\langle\overline{u}''\overline{w}''\rangle}{\partial z}}_{\mathlarger{D2x}} - \underbrace{\frac{\partial\langle\overline{u'u'}\rangle}{\partial x}}_{\mathlarger{R1x}} - \underbrace{\frac{\partial\langle\overline{u'w'}\rangle}{\partial z}}_{\mathlarger{R2x}} - \underbrace{\frac{\partial \langle \overline{p}_t\rangle}{\partial x}}_{\mathlarger{Px}},
    \end{aligned}
\end{align}
where $\{u,v,w\}$ are respectively velocities in the streamwise, spanwise and vertical directions. The variation of all terms in \eqref{eqn:DANSx} with respect to $x$ is illustrated in figure \ref{fig:budget-streamwise} until 150 rotor diameter downstream of the wind farm,  beyond which limited change is observed in the flow quantities. As shown in figure \ref{fig:budget-streamwise} and as expected, the residual term is mostly negligible throughout the entire domain.

\begin{figure}
    \centering
    \includegraphics[width=1\textwidth]{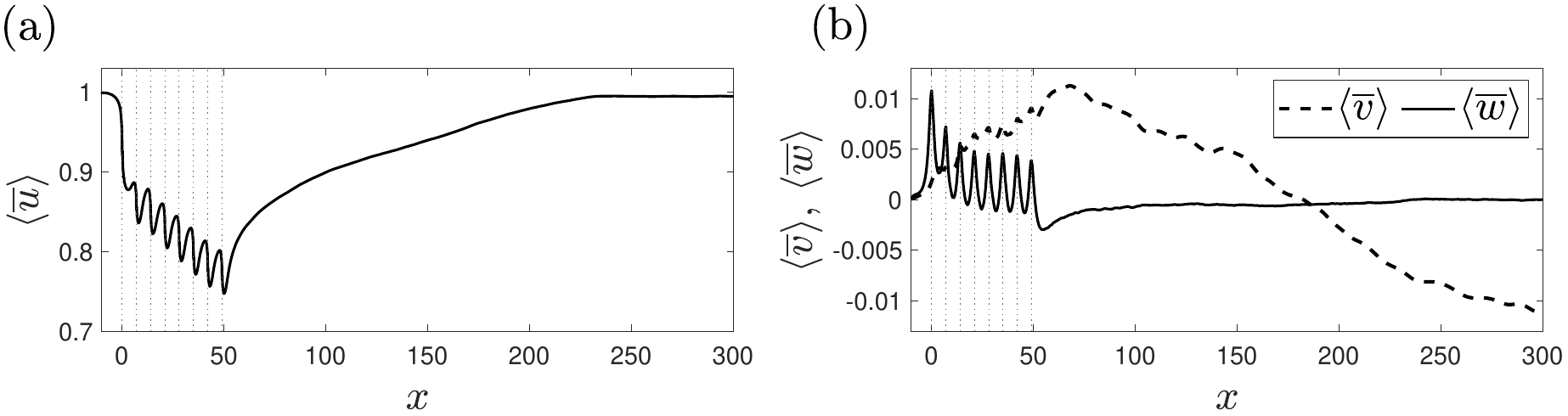}
     \caption{Variation of laterally-averaged (a) streamwise $\langle\bar{u}\rangle$, (b) spanwise $\langle\bar{v}\rangle$ and vertical $\langle\bar{w}\rangle$ velocities with $x$. All variables are non-dimensionalised 
using a selection of $\mathcal{U}_h$ and $D$. The locations of turbine rows are denoted by vertical dotted lines.}
    \label{fig:velProfiles}
\end{figure}

First, we start with the dominant streamwise advection term 
$A1x=\langle\overline{u}\rangle{\partial\langle\overline{u}\rangle}/{\partial x}$ representing the advection of streamwise momentum by the streamwise velocity. A positive value for $A1x$ means wake recovery/flow acceleration, and vice versa. Approaching the farm, $A1x$ becomes negative approximately 8D upstream of the farm. This is explained by the presence of an induction region preceding the farm that is caused by farm-scale blockage effects \citep{bleeg2018blockage}. 
Behind the first row of turbines, we observe that the flow acceleration is still suppressed by farm-scale blockage effects, and $A1x$ continues to be negative until about $3D$, where the maximum velocity deficit occurs (i.e., $A1x=0$). The induction entrance region of the wind farm can be also illustrated by positive vertical velocity $\langle \bar{w}\rangle$ shown in figure \ref{fig:velProfiles}(b).

After the second turbine row, $A1x$ becomes immediately positive indicating that the maximum velocity deficit occurs much closer to the turbine, which is followed by flow acceleration (i.e., wake recovery). By inspection, the profile of vertical Reynolds stress gradient $R2x={\partial\langle\overline{u'w'}\rangle}/{\partial z}$ follows the profile of $A1x$, confirming $R2x$ acts to replenish wake momentum. 
In other words, peak flow acceleration (i.e., maximum $A1x$) is observed, approximately where vertical momentum transport due to turbulence is also maximum.  Note that terms on the right-hand side of \eqref{eqn:DANSx} that are negative in figure \ref{fig:velProfiles} promote wake recovery and vice versa. The greater proportions of turbulent vertical momentum transport in later rows is also evident in figure \ref{fig:budget-streamwise}, occurring due to increased flow shear from greater velocity deficits, as observed in figure \ref{fig:velProfiles}(a). It is worth reminding ourselves that according to \eqref{eqn:DANSx}, the gradient of the Reynolds stress is responsible for wake recovery, as opposed to the common assumption that the mere presence of Reynolds stress promotes wake recovery \citep{van2022brief}. 

Figure \ref{fig:budget-streamwise} also shows that the other advection term $A2x=\langle\overline{w}\rangle {\partial\langle\overline{u}\rangle}/{\partial z}$ has minimal impact on momentum transport within the domain. Moreover, although not discernible in figure \ref{fig:budget-streamwise}, the Coriolis term $Cx=f_c(\langle\overline{v_g}\rangle - \langle\overline{v}\rangle)$ is negative across the domain as expected in the northern hemisphere, but like $A2x$, its value is negligible compared to the dominant terms. 

The normal Reynolds stress gradient $R1x={\partial\langle\overline{u'u'}\rangle}/\partial x$ is illustrative of the rate of change of turbulence level (intensity) with $x$. $R1x$ in figure \ref{fig:budget-streamwise} indicates the turbulence level increases behind each turbine row, peaking about 3-5D downstream (i.e., where $R1x=0$), before decreasing on the approach to the next row. This is in agreement with prior studies observing peak turbulence intensity occurring a few rotor diameters downstream of an individual turbine \citep{Wu2012AtmosphericStudy}. In the wake of the farm, $R1x$ quickly approaches zero as the turbulence decays to its background level.

Figure \ref{fig:budget-streamwise} shows that the turbine pressure gradient $Px={\partial \langle \overline{p}_t\rangle}/{\partial x}$  is significant within the entire farm. 
From actuator disk theory \citep{manwell2010wind}, we know that after the pressure increase upwind of turbines, there is a sudden pressure drop as the turbine extracts energy from the flow (not shown in figure \ref{fig:budget-streamwise}). This is followed by a pressure increase as wake recovery occurs. Figure \ref{fig:budget-streamwise} shows the fast recovery of pressure downwind of each turbine row indicated by positive $P1x$. The value of $P1x$ (i.e., rate of pressure increase) decays with $x$ until it increases again due to the induction region of subsequent rows. It is worth noting that the variation of pressure is often neglected in wake models, but this figure and other recent studies \citep[e.g.,][]{bastankhah2021analytical} highlights the importance of this term. According to figure \ref{fig:budget-streamwise}, within the wind farm, this term is even comparable to other dominant terms (e.g., $A1x$ and $R2x$) in the momentum equation. The term $Px$ however decays quickly in the wake of the wind farm as the pressure approaches its free-stream value.

Some of the most significant terms in figure \ref{fig:budget-streamwise} are the dispersive stress terms which are the product of deviations from the lateral averaging, and described as the tortuous streamlines induced by flow obstacles \citep{Moltchanov2011DispersiveEdge}. Dispersive stresses are correlated to obstacle density. Sparsely populated obstacle fields display increased dispersive stresses, due to greater disparity between flow over the obstacles and the mean flow \citep{Moltchanov2011DispersiveEdge}. For instance, dispersive stresses are expected to be greater in aligned wind farms (shown in figure \ref{fig:budget-streamwise}) than in staggered ones (not shown here) although $D1x$ may be still considerable within a staggered wind farm. 
\begin{figure}
    \centering
    \includegraphics[width = .9\textwidth]{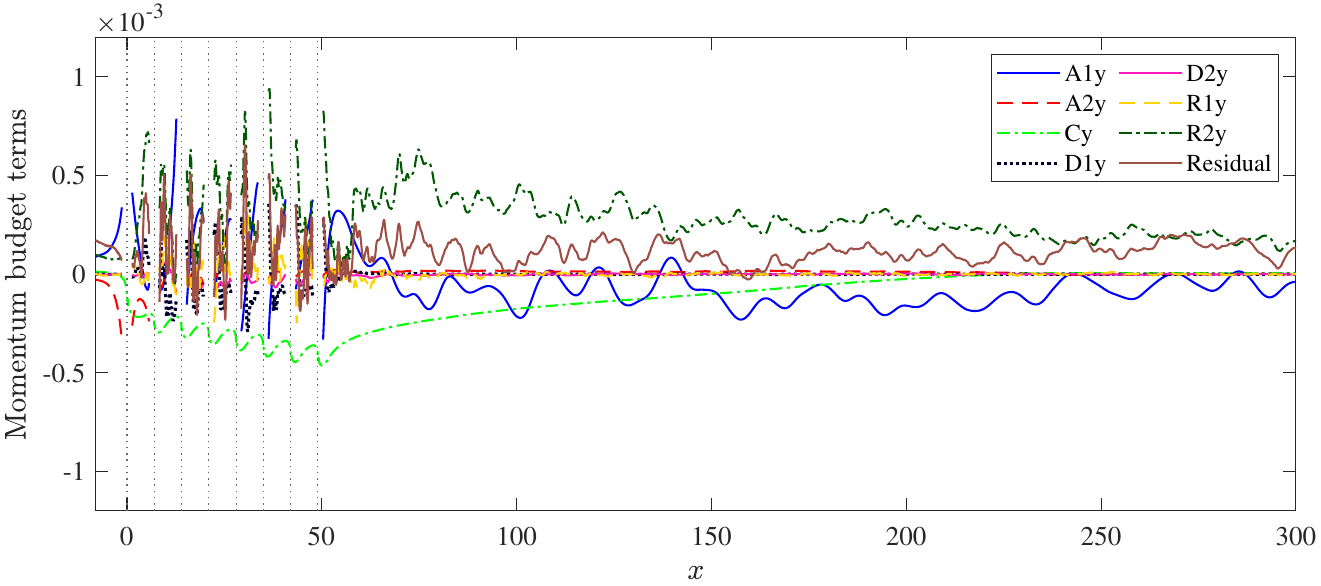}
    \caption{Non-dimensionalised spanwise momentum \eqref{eqn:DANSy} budget at hub height for the Aligned Baseline (A0) case. Locations of turbine rows are denoted by vertical dotted lines. All variables are non-dimensionalised 
using a selection of $\mathcal{U}_h$ and $D$.}
    \label{fig:spanwise-budget}
    \vspace{-0.5cm}
\end{figure}
The term $D1x$ represents the amount of streamwise momentum transport caused by flow inhomogeneity. More precisely, it represents the rate of change in inhomogeneity with respect to the laterally-averaged flow. 
According to figure \ref{fig:budget-streamwise}, $D1x$ is negative within the wind farm indicating the homogeneity of the flow is increasing. This occurs as the wake recovers due to vertical momentum transport into the wake, reducing the magnitude of the spatial fluctuations of streamwise velocity (i.e., $\bar{u}''$). Accordingly, the location of minimum $D1x$ (i.e., maximum rate of approaching homogeneity) is correlated with where maximum vertical momentum transport $R2x$ and maximum wake recovery rate $A1x$ occur. As mixing increases so does the homogeneity, causing the reduction of the magnitude of $D1x$ displayed in figure \ref{fig:budget-streamwise}. Despite its importance within the farm, $D1x$ decays sharply in the wake of the wind farm, where individual turbine wakes merge and form a holistic farm wake. This is discussed in more detail in section \ref{sec:results}. Finally, we investigate the variation of $D2x={\partial\langle\overline{u}''\overline{w}''\rangle}/{\partial z}$. This term essentially quantifies the vertical transfer of streamwise momentum caused directly by the non-uniformity of the time-averaged wind-farm flow field. Figure \ref{fig:budget-streamwise} shows that this term is clearly smaller than the other dispersive term $D1x$. It is also interesting to note that this term is mainly positive within the wind farm. This indicates that $D2x$ acts against wake recovery, in contrast to its turbulent counterpart $R2x$.

\subsection{Spanwise momentum equation}\label{sec:budget_analysis_spanwise}
Even though the terms of the spanwise momentum equation are of smaller magnitude than their streamwise counterparts, they are examined here due to their importance to the wake's trajectory. Writing \eqref{eqn:LANS} in the spanwise direction (i.e., $i=2$) yields
\begin{align} \label{eqn:DANSy}
\begin{aligned}
    \underbrace{\langle\overline{u}\rangle\frac{\partial\langle\overline{v}\rangle}{\partial x}}_{\mathlarger{A1y}}  + \underbrace{\langle\overline{w}\rangle\frac{\partial\langle\overline{v}\rangle}{\partial z}}_{\mathlarger{A2y}} = -\underbrace{f_c(\langle\overline{u}\rangle-\langle\overline{u_g}\rangle)\vphantom{\frac{\partial}{\partial}}}_{\mathlarger{Cy}}  - \underbrace{\frac{\partial \langle\overline{v}''\overline{u}''\rangle}{\partial x}}_{\mathlarger{D1y}}  - \underbrace{\frac{\partial\langle\overline{v}''\overline{w}''\rangle}{\partial z}}_{\mathlarger{D2y}} - \underbrace{\frac{\partial\langle\overline{v'u'}\rangle}{\partial x}}_{\mathlarger{R1y}} - \underbrace{\frac{\partial\langle\overline{v'w'}\rangle}{\partial z}}_{\mathlarger{R2y}}.
    \end{aligned}
\end{align}
Variations of terms in \eqref{eqn:DANSy} with the streamwise distance $x$ are shown in figure \ref{fig:spanwise-budget}. Due to their small values, the terms in \eqref{eqn:DANSy} are more prone to numerical errors, which may explain why their variations are more oscillatory and less smooth compared to their streamwise counterparts. 

First, we start with the Coriolis term $Cy=f_c(\langle\overline{u}\rangle-\langle\overline{u_g}\rangle)$. The dominance of this term varies across the domain. From figure \ref{fig:spanwise-budget}, $Cy$ is dominantly negative within the farm due to the streamwise flow deceleration, which according to \eqref{eqn:DANSy} causes positive $A1y=\langle\overline{u}\rangle{\partial\langle\overline{v}\rangle}/{\partial x}$ and thereby positive $\langle \bar{v} \rangle$, as observed in figure \ref{fig:velProfiles}(b). This deflects the wake to the left, which can be described as an anticlockwise flow rotation viewed from the top. In the wind-farm wake, figure \ref{fig:spanwise-budget} however shows that with flow acceleration the strength of the Coriolis force decays. This is where $R2y={\partial\langle\overline{v'w'}\rangle}/{\partial z}$ that is the vertical turbulent entrainment of veered momentum from above becomes more dominant. Consequently, this changes the sign of the advection term $A1y$ to negative as shown in figure \ref{fig:spanwise-budget}, and therefore the far wake starts deflecting to the right (i.e., clockwise wake rotation viewed from top). In other words, the term $A2y$ turns the wind at the hub height towards the wind direction at higher altitudes. This ultimately leads to a negative spanwise velocity as depicted in figure \ref{fig:velProfiles}(b). This interesting phenomenon is discussed in more detail in section \ref{sec:results}. It is also noteworthy that the magnitude of the second advection term $A2y$, the dispersive stress terms $D1y$ and $D2y$ and also $R1y$ are small, especially in the farm wake and can be neglected.

\subsection{Approximate form of momentum equations}
From the analysis in section \ref{sec:budget_analysis_streamwise}, amongst others the dispersive stress ($D1x$) and pressure ($Px$) terms are evidently not negligible in the streamwise momentum equation, at least within the farm region. However, despite the evident importance of these two terms, the summation of the four terms $D1x$, $D2x$, $R1x$, and $Px$—which are challenging to model—is rather small. This is illustrated by the dashed light green color in figure \ref{fig:budget-streamwise}. The combined value of these terms is negligible in the wake of the farm. Within the farm, the combined value is not negligible but smaller than the individual dispersive $D1x$ and pressure $Px$ terms. The term $D1x$ is negative, while $Px$ is positive; therefore, to some extent, they cancel each other out.
For simplicity, we thus omit these terms from our model developed in section \ref{sec:model_development}. Moreover, as discussed in section \ref{sec:budget_analysis_spanwise}, it is apparent that in the spanwise direction the dominant terms are (i) the spanwise momentum advection by the streamwise velocity $A1y$, (ii) the Coriolis term $Cy$ and (iii) the vertical turbulent transport of veering wind $R2y$. Therefore, the approximate forms of the momentum equations including turbine forcing can be written as
\begin{align}\label{eqn:DANS_simplified}
\begin{split}
       \langle \overline{u}\rangle \frac{\partial \langle \overline{u}\rangle}{\partial x} &\approx 
    -f_c (\langle \overline{v_{\mathit{g}}}\rangle  -\langle \overline{v}\rangle)\vphantom{\frac{\rangle\partial}{\partial}}  
     - \frac{\partial \langle \overline{u'w'}\rangle}{\partial z}
   -\langle  \bar{f}_{x} \rangle,\\ 
       \langle \overline{u}\rangle \frac{\partial \langle \overline{v}\rangle}{\partial x} &\approx 
    +f_c (\langle \overline{u_{\mathit{g}}}\rangle  -\langle \overline{u}\rangle)\vphantom{\frac{\rangle\partial}{\partial}}  
     - \frac{\partial \langle \overline{v'w'}\rangle}{\partial z}
   -\langle  \bar{f}_{y} \rangle.
   \end{split}
\end{align}
In the following section, we simplify \eqref{eqn:DANS_simplified} to develop a system of ordinary differential equations (ODEs) that can be solved mathematically for a semi-infinite wind farm.

\section{Derivation of physics-based fast-running farm wake model}\label{sec:model_development}
\subsection{Definition of semi-infinite wind farm}\label{sec:wind_farm_definition}
\begin{figure}
    \centering
    \includegraphics[width=.8\textwidth]{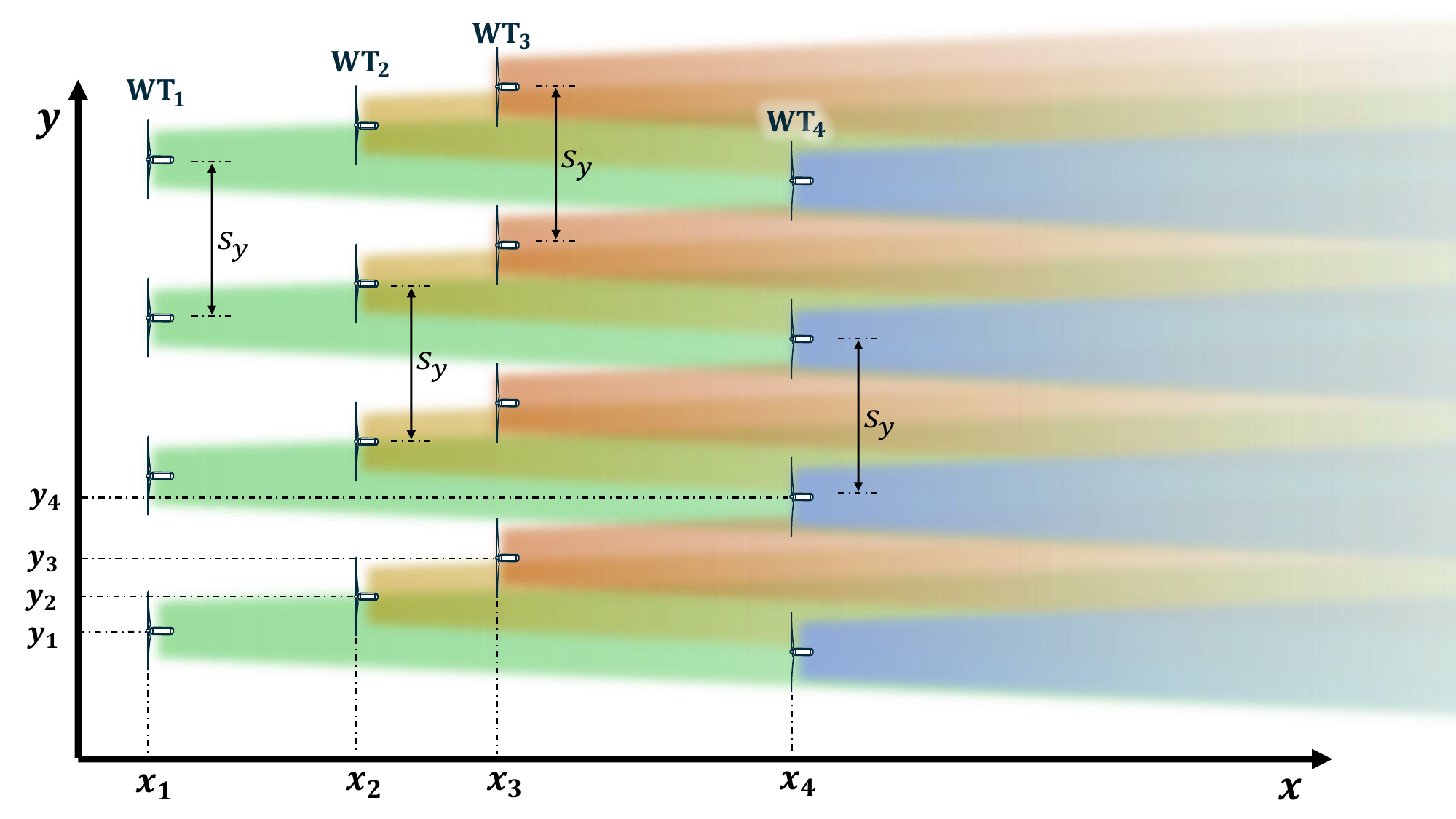}
    \caption{Schematic of a semi-infinite wind farm. Turbines in row $n$ are denoted by WT$_n$. The lateral spacing $s_y$ between turbines is assumed to be constant for the whole wind farm. However, the streamwise spacing can be variable, and moreover turbine rows may be laterally shifted with respect to each other.}
    \label{fig:schematic_wf}
\end{figure}

We start by assuming a semi-infinite wind farm which is infinite in the lateral direction but has a finite length in the streamwise direction. A right-handed Cartesian coordinate system $\{x,y,z\}$ aligned with the incoming wind at the turbine hub height $z_h$ is adopted such that $x$ is in the direction of the incoming wind, $y$ represents the horizontal direction normal to $x$, and $z$ measures the height from the ground. The total number of wind turbine rows is denoted by $N$. Wind turbines in the n$^{th}$ row are abbreviated to WT$_n$s, where the subscript $n=\{1,2,...,N\}$ shows the row number labelled based on the streamwise position (i.e., ranging from $n=1$ for the first row to $n=N$ for the last row). WT$_n$s are assumed to have the same values of thrust coefficient $C_{T,n}$ and yaw angle $\gamma_n$, which may however be different from $C_{T,m}$ and $\gamma_m$ if $n\neq m$. In other words, turbines in different rows may have different operating conditions. While the lateral spacing $s_y$ between turbines is assumed to be the same for all rows, the streamwise spacing between consecutive rows may be variable. The arbitrary streamwise positions of turbine rows is quite advantageous, as for instance the developed model can be even applied to a cluster of wind farms at once (not done in this study).  Furthermore, turbine rows may have lateral offset with respect to each other as shown in figure \ref{fig:schematic_wf}. The lateral position of WT$_n$s is $y_n+ks_y$ where $k=-\infty\dots\infty$.  
\subsection{turbine force}
The normalised aerodynamic force $\bar{f}$ exerted by wind turbines is given by
\begin{align}
\begin{split}
    \bar{f}_x &=-\sum_{n=1}^N\sum_{k=-\infty}^\infty\frac{1}{2} C_{T,n}\bar{u}_{h,n}^2\cos(\gamma_n)\delta\left(x-x_n\right)H\left(0.5^2-\left[\left(y-y_n-ks_y\right)^2+\left(z-z_h\right)^2\right]\right),\\
    \bar{f}_y&=-\sum_{n=1}^N\sum_{k=-\infty}^\infty\frac{1}{2} C_{T,n}\bar{u}_{h,n}^2\sin(\gamma_n)\delta\left(x-x_n\right)H\left(0.5^2-\left[\left(y-y_n-ks_y\right)^2+\left(z-z_h\right)^2\right]\right),
    \end{split}
\end{align}
where $\bar{u}_{h,n}$ is the local incoming hub-height velocity for WT$_n$s. In other words, it is the time-averaged velocity at the location of WT$_n$s' rotor centre (i.e., $(x,y,z)=(x_n,y_n,z_h)$) in the absence of WT$_n$s. $\delta(x)$ is the Dirac delta function, and $H(x)$ is the Heaviside step function, which is defined as $H(x)=0$ for $x\leq0$ and $H(x)=1$ for $x>0$. We then apply the lateral averaging discussed in section \ref{sec:LA-RA-NSE} to obtain laterally-averaged turbine forces at the hub height $z=z_h$ as follows
\begin{align}
\begin{split}\label{eqs:turbine_forcing}
    \langle\bar{f}_x\rangle &=-\frac{1}{2s_y}\sum_{n=1}^N C_{T,n}\bar{u}_{h,n}^2\cos(\gamma_n)\delta\left(x-x_n\right),\\
    \langle\bar{f}_y\rangle&=-\frac{1}{2s_y}\sum_{n=1}^N C_{T,n}\bar{u}_{h,n}^2\sin(\gamma_n)\delta\left(x-x_n\right).
    \end{split}
\end{align}
\subsection{Simplifying Reynolds shear stress terms in \eqref{eqn:DANS_simplified}}\label{sec:shear_stress}
\begin{figure}
    \centering
        \begin{overpic}[width=1\textwidth]{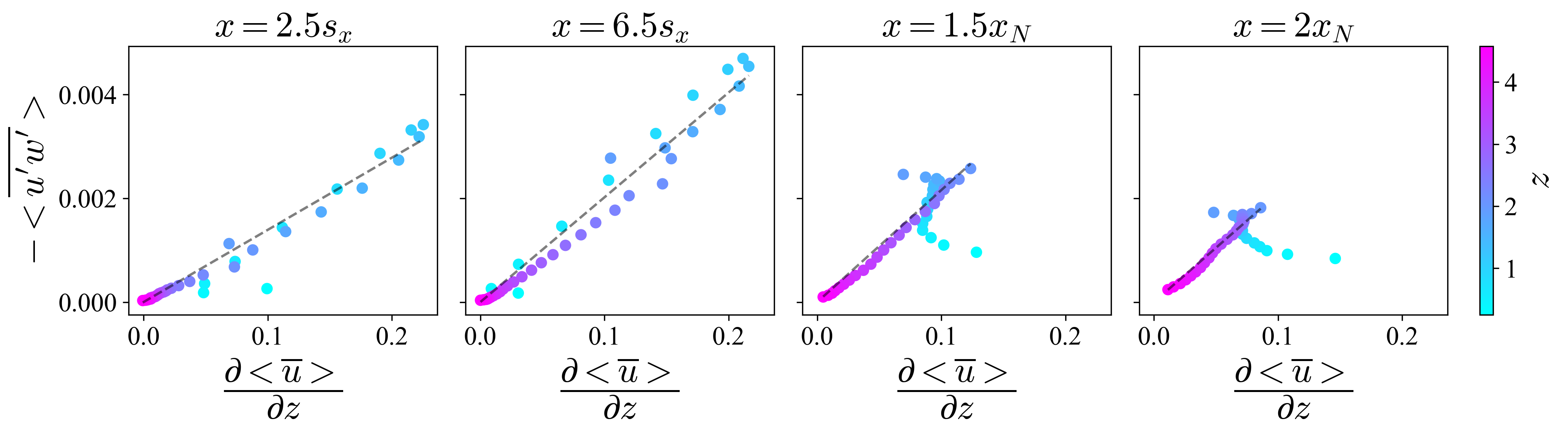}
           \put(9,22){(a)} 
  \put(30.5,22){(b)}
   \put(52,22){(c)}
    \put(74,22){(d)}
    \end{overpic}
    \caption{Distribution of $-\langle \overline{u'w'}\rangle$ against $\partial \langle \bar{u}\rangle/\partial z$ at different heights for four different streamwise positions in the A0 case, where (a-b) are in the farm region, and (c-d) are in the wake region, and $x_N=7s_x$ is the farm length.}
    \label{fig:nu_t_validity}
\end{figure}
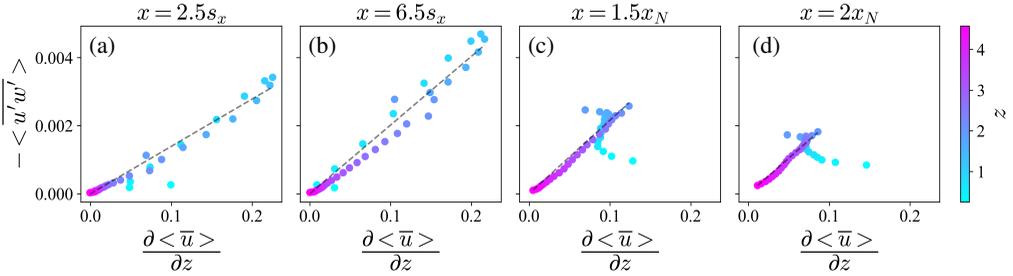

The Boussinesq eddy-viscosity hypothesis has been used in previous studies \citep{luzzatto2018,belcher-Hunt2003} to model spatially-averaged Reynolds stresses based on spatially-averaged velocity gradients, 
\begin{equation} \label{eqn:bous}
    \langle \overline{u_i'u_j'}\rangle = -2\nu_t \bigg[ \underbrace{\frac{1}{2}\bigg(\frac{\partial \langle \overline{u_i}\rangle}{\partial x_j} + \frac{\partial \langle \overline{u_j}\rangle}{\partial x_i}\bigg)}_{\mathlarger{S_{ij}}}\bigg], \;\; (\textrm{for}\;\;i\neq j)
\end{equation}
where $S_{ij}$ is the dimensionless rate of strain tensor, and $\nu_t$ is the turbulent viscosity non-dimensionalised by $\mathcal{U}_hD$. For $i=1$ and $j\neq 1$, $\partial u_i/\partial x_j$ is an order of magnitude larger than $\partial u_j/\partial x_i$ \citep{tennekes1972first}, so one can write
\begin{equation} \label{eqn:vert_bous}
  \frac{ \partial \langle \overline{u'w'}\rangle}{ \partial z} \approx -\nu_t \frac{\partial^2 \langle \overline{u}\rangle}{\partial z^2},\;\;\;\frac{\partial \langle \overline{v'w'}\rangle}{\partial z} \approx -\nu_t \frac{\partial^2 \langle \overline{v}\rangle}{\partial z^2}.
\end{equation}
We first assess the validity of the turbulent-viscosity hypothesis using our LES data. To do so, at a given streamwise position, we examine whether $-\langle \overline{u'w'}\rangle$ is linearly proportional to $\partial \langle \bar{u}\rangle/\partial z$, according to \eqref{eqn:bous}. Figure \ref{fig:nu_t_validity} shows values of $-\langle \overline{u'w'}\rangle$ and $\partial \langle \bar{u}\rangle/\partial z$ at different heights and streamwise positions. Results are shown for the A0 case, but they look qualitatively similar in other cases (not shown here). Data are plotted for a range of $z=0.25$ (shown in light blue) to $z=4.57$ (shown in magenta). Results in figure \ref{fig:nu_t_validity} are shown for four streamwise locations, with the first two in the farm region and the other two in the wake region: (a) between WT$_3$ and WT$_4$ ($x=2.5s_x$), (b) between WT$_7$ and WT$_8$ ($x=6.5s_x$), (c) half of the farm length downstream in the wake ($x=1.5x_N$), where the farm length is $x_N$, (d) an entire farm length downstream in the wake ($x=2x_N$). Lines are fitted to the data at heights above the hub height. Figure \ref{fig:nu_t_validity}(a-b) shows that within the farm region, the eddy-viscosity assumption seems to be a valid approximation for the entire vertical domain plotted in the figure, except for the region close to the ground. In the farm wake, however, this assumption is only valid at upper heights as shown in figure \ref{fig:nu_t_validity}(c-d). Given that the height at which the eddy viscosity assumption appears reasonable in the farm wake seems to grow with downstream distance, one possible explanation for this could involve the development of the secondary internal boundary layer (IBL) downwind of the wind farm due to the transition from rough to smooth terrain (see section \ref{sec:turbulent_viscosity} for more discussion on IBLs). However, further research is required to fully elucidate this phenomenon. Nevertheless, even in the farm wake, the turbulent-viscosity hypothesis is still valid at upper heights, where the top-down transport of energy by Reynolds shear stresses occurs, so we use the turbulent-viscosity assumption in this work to simplify the laterally-averaged momentum equations.

The turbulent viscosity $\nu_t$ is then decomposed into two parts: one that is due to the ambient atmospheric turbulence denoted by $\nu_{t,0}$ and one shown by $\nu_{t,f}$ corresponding to the turbulence added by the wind farm, and it varies by $x$. In other words, 
\begin{equation}
\nu_t(x)=\nu_{t,0}+\nu_{t,f}(x).
\end{equation}
Specifying values of the ambient turbulent viscosity $\nu_{t,0}$ and the farm turbulent viscosity $\nu_{t,f}$ is deferred to section \ref{sec:turbulent_viscosity}.
\subsection{Mathematical solution of velocity deficit equations}
Now, we develop and solve equations for the variation of laterally-averaged velocity deficit in both streamwise $U_d$ and spanwise directions $V_d$, defined as 
\begin{align}
U_d(x)&=U_0- U(x), \\
V_d(x)&=V_0-V(x).
\end{align}
In the above equation and hereafter, $U(x)=\langle\bar{u}\rangle(x,z=z_h)$, $V(x)=\langle\bar{v}\rangle(x,z=z_h)$. The subscript $0$ denotes the flow in the absence of the whole wind farm. Let us recall that the coordinate system is defined based on the incoming wind direction at the hub height (see section \ref{sec:wind_farm_definition}) and all velocities in this work are non-dimensionalised by $\mathcal{U}_h$, so $U_0=1$ and $V_0=0$. The latter means that $V_d=-V$ in this work. For generality, however, we write equations for $V_d$ so the model can still be used for a different coordinate system where $V_0\neq 0$.

The magnitude of the local velocity deficit caused by each individual turbine can be significant especially in the near-wake region \citep[e.g.,][]{zhang2012near,Bastankhah2017POF}. However, laterally-averaged values of the velocity deficit are fairly small, as will be later shown in section \ref{sec:results}. Therefore, we linearise \eqref{eqn:DANS_simplified}  by replacing $\langle\bar{u}\rangle$ in the advection term with the incoming velocity $U_0=1$. Moreover, using the turbulent-viscosity hypothesis discussed in section \ref{sec:shear_stress} to simplify \eqref{eqn:DANS_simplified}, we obtain
\begin{align}\label{eqn:DANS_turb_visc}
\begin{split}
    \frac{\partial U}{\partial x} &\approx 
    -f_c (V_{\mathit{g}}  -V)\vphantom{\frac{\rangle\partial}{\partial}}  
     +\nu_t\frac{\partial^2 U}{\partial z^2}
   -\langle  \bar{f}_{x} \rangle,\\
      \frac{\partial V}{\partial x} &\approx 
    +f_c (U_{\mathit{g}}  -U)
     +\nu_t \frac{\partial^2 V}{\partial z^2}
     -\langle  \bar{f}_{y} \rangle.
\end{split}
\end{align}
In the absence of the wind farm, \eqref{eqn:DANS_turb_visc} is simplified to
\begin{align}\label{eqn:DANS_ekman}
\begin{split}
   0 &\approx 
    -f_c (V_{\mathit{g}}  -V_0)\vphantom{\frac{\rangle\partial}{\partial}}  
     +\nu_{t,0} \frac{\partial^2 U_0}{\partial z^2},\\ 
      0 &\approx 
    +f_c (U_{\mathit{g}}  -U_0)
     +\nu_{t,0} \frac{\partial^2 V_0}{\partial z^2}.
     \end{split}
\end{align}
It is worth noting that \eqref{eqn:DANS_ekman} was solved in \cite{ekman1905} for different altitudes to describe the well-known Ekman spiral. If we subtract \eqref{eqn:DANS_turb_visc} from \eqref{eqn:DANS_ekman} and also use the dimensional analysis of $\partial^2 U_d/\partial z^2\propto U_d/l^2$ where $l\approx 1$ (i.e., length scale comparable to rotor diameter), we obtain
\begin{align}
\begin{split}\label{eqs:velocity_defict}
    \frac{\textrm{d} U_d}{\textrm{d} x} \approx &+
    \underbrace{f_cV_d\vphantom{\frac{a}{b}}}_{\text{\small Coriolis}}        -\underbrace{c_1 \nu_t U_d \vphantom{\frac{a}{b}}}_{\text{\small recovery}}    +\underbrace{\langle  \bar{f}_{x}\rangle \vphantom{\frac{a}{b}}}_{\text{\small turbine forcing}}+ \underbrace{C_{x}\vphantom{\frac{a}{b}}}_{\text{\small shear}},\\
      \frac{\textrm{d} V_d}{\textrm{d} x} \approx &-
    \underbrace{f_c U_d\vphantom{\frac{a}{b}}}_{\text{\small Coriolis}}        -\underbrace{c_1 \nu_t V_d \vphantom{\frac{V_d}{Ro}}}_{\text{\small recovery}}    +\underbrace{\langle  \bar{f}_{y}\rangle \vphantom{\frac{V_d}{Ro}}}_{\text{\small turbine forcing}}+ \underbrace{C_{y}\vphantom{\frac{V_d}{Ro}}}_{\text{\small veer}},\\
\end{split}
\end{align}
where $c_1$ is a constant. The term  $C_{x}=-\nu_{t,f} \partial^2 U_0/\partial z^2$ is related to the rate of incoming shear, so it is called \textit{shear} in \eqref{eqs:velocity_defict}. Likewise, $C_{y}=-\nu_{t,f} \partial^2 V_0/\partial z^2$ is related to the rate of incoming veer, and it is called \textit{veer} in \eqref{eqs:velocity_defict}. Both $C_x$ and $C_y$ also depend on the amount of turbulence generated by the wind farm through $\nu_{t,f}$. Values of $C_x$ and $C_y$ are determined as a function of atmospheric and farm conditions later in section \ref{sec:Cx and Cy}, and for now they are assumed to be known.

If we insert \eqref{eqs:turbine_forcing} into \eqref{eqs:velocity_defict}, the below mathematical solution for this system of ODEs can be obtained. The solution written in \eqref{eqs:final_form} is exact for a constant $\nu_{t,f}$. For a well-defined function of $\nu_{t,f}(x)$ with an arbitrary distribution, this provides an approximate solution with insignificant error with respect to the exact numerical solution (not shown here) of the ODE system.
\begin{align}\label{eqs:final_form}
    \begin{split}
        U_d(x)\approx &\frac{C_x}{c_1\nu_t }\left(1-\textrm{e}^{-c_1\int_{0}^{x} \nu_t\textrm{d}x}\right)H(x)+\sum_{n=1}^N U_{d,n}(x),\\
         V_d(x)\approx &\frac{C_y}{c_1\nu_t }\left(1-\textrm{e}^{-c_1\int_{0}^{x} \nu_t\textrm{d}x}\right)H(x)+\sum_{n=1}^N V_{d,n}(x),
    \end{split}
\end{align}
where $U_{d,n}(x)$ and $V_{d,n}(x)$ are, respectively, values of the laterally-averaged streamwise and spanwise velocity deficit caused by WT$_n$s at $x$, and they are given by:
\begin{align}\label{eqs:final_form_Udn}
    \begin{split}
        U_{d,n}(x)=\frac{C_{T,n}}{2s_y}\left(1-\eta_n U_d(x_n)\right)^2\textrm{e}^{-c_1\int_{x_n}^{x} \nu_t\textrm{d}x}\cos\left(\gamma_n-f_c(x-x_n)\right)H(x-x_n),\\
         V_{d,n}(x)=\frac{C_{T,n}}{2s_y}\left(1-\eta_n U_d(x_n)\right)^2\textrm{e}^{-c_1\int_{x_n}^{x} \nu_t\textrm{d}x}\sin\left(\gamma_n-f_c(x-x_n)\right)H(x-x_n).
    \end{split}
\end{align}
In \eqref{eqs:final_form_Udn}, $\bar{u}_{h,n}$ is replaced with
\begin{equation}
    \bar{u}_{h,n}=1-\eta_n U_d(x_n),
\end{equation}
where $U_d(x_n)=U_d(x=x_n)$, and we call $\eta$ the local deficit coefficient, which is defined as the ratio of the local velocity deficit experienced by wind turbines to the laterally-averaged velocity deficit. This coefficient depends on the farm layout configuration, and it is determined by \eqref{eq:alpha} developed later in section \ref{sec:local_hub_height_ve}. To compute the ambient turbulent viscosity $\nu_{t,0}$ and farm turbulent viscosity $\nu_{t,f}$, \eqref{eq:nut0} and \eqref{eq:nutf} developed in section \ref{sec:turbulent_viscosity} are respectively used. Finally, values of $C_x$ and $C_y$ are estimated based on \eqref{eq:cx-cy-final} in section \ref{sec:Cx and Cy}. Once values of $C_x$, $C_y$, $\nu_{t,0}$ and $\nu_{t,f}$ are all known, a forward marching scheme in the streamwise direction is implemented using \eqref{eqs:final_form} to find the evolution of $U_d$ and $V_d$ with $x$. The forward marching scheme is stable and not sensitive to the streamwise resolution, which was tested (not shown here) for a range of values from $0.01D$ to $1D$. The solution in \eqref{eqs:final_form} is versatile as it can also be used for different distributions of $\nu_{t}$, $C_x$ and $C_y$ that can potentially be developed in future studies.


\subsection{Estimation of turbulent viscosity $\nu_t=\nu_{t,0}+\nu_{t,f}$}\label{sec:turbulent_viscosity}
In general, turbulent viscosity can be written as a product of a turbulence velocity scale $\hat{u}$ and a turbulence length scale (i.e., mixing length) $\hat{l}$ \citep{tennekes1972first}. Therefore, to estimate the ambient turbulent viscosity $\nu_{t,0}$ and the farm turbulent viscosity $\nu_{t,f}$, one needs to specify suitable values of $\hat{u}$ and $\hat{l}$ for each term. 

For the ambient turbulent viscosity $\nu_{t,0}$, according to Prandtl's mixing-length hypothesis \citep{tennekes1972first}, we assume $\hat{u}_0=u_*$ and $\hat{l}_0=\kappa z_h$ where $\kappa\approx 0.41$ is the von-K\'{a}rma\'{n} constant. Therefore, the ambient turbulent viscosity is given by
\begin{equation}\label{eq:nut0}
    \nu_{t,0}=\kappa u_*z_h.
\end{equation}
It is worth noting that the assumption of $\hat{l}_0=\kappa z_h$ is expected to be valid only in the log-law region of the ABL \citep{pope2000}. For our simulations the top of the rotor disk is at a location $z/H = 0.22$ which is seemingly right at the outer limit of this assumption. More sophisticated models for the mixing length in ABLs can be used instead \citep[e.g.,][]{van2020rossby} but for simplicity we retain the simple log-law relationship for $\hat{l}_0$.

For the farm turbulent viscosity $\nu_{t,f}$, we use the height of the internal boundary layer (IBL) that grows above the wind farm as the turbulence length scale. Several studies have shown that the IBL grows above wind farms following the classical Elliott's $x^{0.8}$ power law \citep[e.g.,][]{allaerts2017,Wu2017FlowFarms}. According to \cite{wood1982}, the thickness of the IBL, $\delta$, due to a smooth to rough transition is given by
\begin{equation}\label{eqs:surface_roughness}
    \frac{\delta}{z_{0,f}}=0.28\left(\frac{x}{z_{0,f}}\right)^{0.8},
\end{equation}
where $z_{0,f}$ is the equivalent roughness length of the wind farm. It is worth noting that the IBL may become separated from the ground surface as a new IBL starts developing due to the rough to smooth transition downwind of the wind farm \citep{oke1976}. For simplicity, we use the maximum height of the first IBL as the turbulence length scale and do not account for the second IBL development. In order to use \eqref{eqs:surface_roughness}, one needs to estimate the value of $z_{0,f}$ for the wind farm. For an infinite wind farm, several models have been already proposed in the literature to estimate $z_{0,f}$ \citep[e.g.,][]{frandsen1992, calaf2010large, abkar2013, yang2012}. The one suggested by \cite{frandsen1992} for an infinite wind farm with uniformly distributed wind turbines states 
\begin{equation}\label{eq:z0-frandsen}
    z_{0,f}=z_h\exp\left(-\frac{\kappa}{\sqrt{\frac{1}{2}c_{ft}+\left[\frac{\kappa}{\ln(z_h/z_{0,0})}\right]^2}}\right),
\end{equation}
where $c_{ft}=\pi C_T/(4s_xs_y)$, $s_x$ is the normalised streamwise spacing between turbine rows, and as a reminder $z_{0,0}$ is the normalised surface roughness length in the absence of the wind farm. To maintain simplicity, we use the same relationship to estimate $z_{0,f}$ for the semi-infinite wind farm. To compute $c_{ft}$, we use the average streamwise inter-turbine spacing for $s_x$ (i.e., $s_x=\sum_{n=1}^{N-1}(x_{n+1}-x_{n})/(N-1)$), and the average value of thrust coefficient for $C_T$ (i.e., $C_T=\sum_{n=1}^{N}C_{T,n}/N$). 

\begin{figure}
\centering
\includegraphics[width=.85\linewidth]{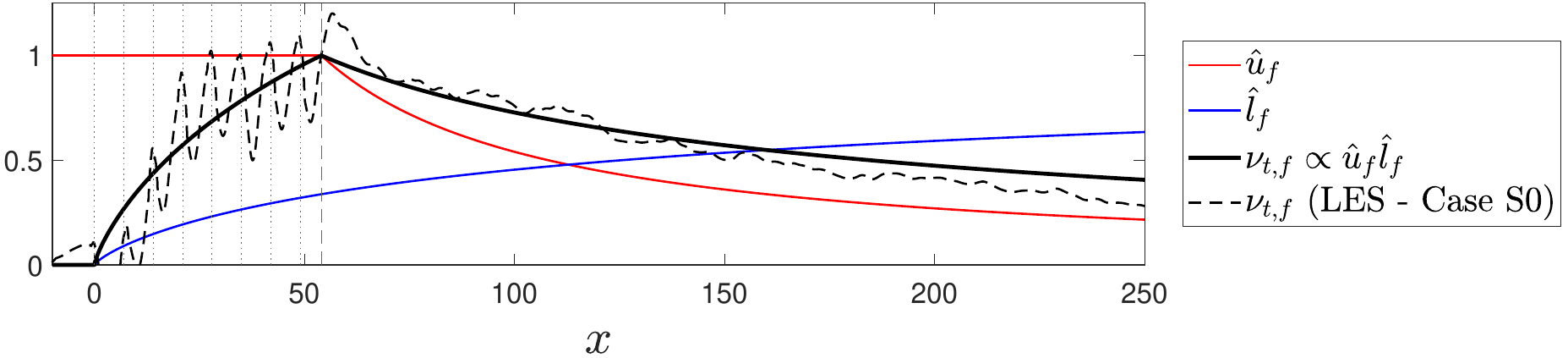}
\caption{Variation of velocity scale $\hat{u}_f$, length scale $\hat{l}_f$ and farm turbulent viscosity $\nu_{t,f}$ based on the modelling approached elaborated in section \ref{sec:turbulent_viscosity}. Variation of $\nu_{t,f}$ for the LES data (case S0) is also shown. In the figure, $\hat{u}_f$ is normalised by $\sqrt{c_{ft}}$, $\nu_{t,f}$ is normalised by $\nu_{t,f}(x=x_1+L_f)$, and $\hat{l}_f$ is normalised by $H$. Vertical dotted lines show the location of turbine rows, and the vertical dashed line shows $x_1+L_f$ where the turbulent viscosity $\nu_{t,f}$ is maximum.} \label{fig:turbulent_viscosity}
\end{figure}

One can use \eqref{eqs:surface_roughness} in conjunction with \eqref{eq:z0-frandsen} to estimate the thickness of the IBL over the wind farm. This relationship is however only valid as long as $\delta$ is smaller than the ABL thickness $H$ \citep{wood1982}. It is known that the IBL growth is capped by the thermal inversion at the top of the ABL \citep{oke1976}, especially if the inversion layer has strong free-atmosphere stratification. In these cases, the inversion layer may act as a ``lid'' on the top of the ABL and hinders the growth of IBL. We therefore artificially limit the growth of the turbulence length scale $\hat{l}_f$ using the below relationship,
\begin{equation}\label{eq:turb_length-scale}
    \hat{l}_f=\frac{\delta}{1+\delta/H}.
\end{equation}
For small values of $x$, $\hat{l}_f\approx\delta$, while $\hat{l}_f\to H$ as the value of $x\to\infty$. Variation of $\hat{l}_f$ with $x$ based on \eqref{eq:turb_length-scale} is shown in figure \ref{fig:turbulent_viscosity}. For simplicity, we here assume that the value of $H$ is constant. It is however important to note that depending on the size of the wind farm and the level of thermal stratification in the inversion layer, the IBL may lead to the growth of the entire ABL, so in reality $H$ might change with $x$ \citep{allaerts2017,Wu2017FlowFarms}.

Next, we determine the turbulence velocity scale $\hat{u}_f$. Within the wind farm, turbulence is mainly generated due to the shear caused by turbine forcing. Therefore, inspired by \citep{calaf2010large}, we use $\sqrt{c_{ft}}U_0$, where $U_0=1$, to estimate the turbulence velocity scale $\hat{u}_f$ within the wind farm. The turbulence generation is expected to peak at some distances downstream of the wind farm, which is then followed by a decay in turbulence generation due to wake recovery and reduction of flow shear \citep{stieren2022impact}. The generated turbulence in the turbine wake usually peaks at around $5$ rotor diameter downstream \citep{chamorro2009roughness-transition}. The constant turbulence velocity scale $\hat{u}_f$ caused by turbine forcing is therefore assumed to be extended from $x_1$ to $x_N+5$ as shown in figure \ref{fig:turbulent_viscosity}. Further downstream, $\hat{u}_f$ starts to decay due to the wake recovery. The wake of a semi-infinite wind farm can be modelled as a two-dimensional wake of a canopy of roughness elements in a turbulent boundary layer. According to \cite{belcher-Hunt2003}, the velocity scale defined based on the maximum velocity deficit in this type of wake flows decays with $x^{-1}$. So in summary, we model the turbulence velocity scale $\hat{u}_f$ as
\begin{equation}
  \hat{u}_f =
    \begin{cases}
    0 & \text{if $x<x_1$}\\
      \sqrt{c_{ft}} & \text{if $x_1<x<x_1+L_f$}\\
      \sqrt{c_{ft}}\left(\frac{x-x_1}{L_f}\right)^{-1} & \text{if $x>x_1+L_f$}
    \end{cases}       
\end{equation}
where $L_f=(x_N-x_1)+5$. Variation of $\hat{u}_f$ and $\nu_{t,f}$ are shown in figure \ref{fig:turbulent_viscosity}, where the farm turbulent viscosity $\nu_{t,f}$ is given by
\begin{equation} \label{eq:nutf}
    \nu_{t,f}=c_2\hat{u}_f\hat{l}_f,
\end{equation}
with $c_2$ a constant. As can be seen in figure \ref{fig:turbulent_viscosity}, the farm turbulent viscosity increases within the farm until it reaches its maximum value five rotor diameter downstream of the wind farm. It then decreases further downstream until it eventually becomes zero very far downstream.  In general, this is in fairly good agreement with our LES data. As an example, the variation of $\nu_{t,f}$ for Staggered Baseline (S0) case is shown in figure \ref{fig:turbulent_viscosity}. Fairly similar variations of turbulent viscosity have been also reported in the literature for the wake of a single wind turbine \citep{scott2023}.

\subsection{Estimation of local deficit coefficient $\eta_n$}\label{sec:local_hub_height_ve}
As discussed earlier, $\bar{u}_{h,n}=1-\eta_nU_d(x=x_n)$, where $\eta_n$ is the local deficit coefficient for WT$_n$s. A consequence of linearising the momentum equation \eqref{eqn:DANS_turb_visc} is that wake effects are linearly superposed \citep{bastankhah2021analytical}. Therefore, one can write
\begin{equation}\label{eq:app:hub_height_velocity}
  \eta_{n} =\frac{\sum_{m=1}^{n-1} \bar{u}_{d,m}(x_n,y_n,z_h)}{\sum_{m=1}^{n-1} U_{d,m}(x_n)},
  \end{equation}
where $\bar{u}_{d,m}(x,y,z)$ is the local velocity deficit at $(x,y,z)$ caused by WT$_m$s, and $U_{d,m}(x)$ is already defined in \eqref{eqs:final_form_Udn}. 
To find $\bar{u}_{d,mn}$, one can use a Gaussian distribution $C\textrm{exp}(-r^2/2\sigma^2)$ to express the velocity deficit caused by each turbine, where $C$ is the wake-centre velocity deficit, $r$ is the distance from the wake centre, and $\sigma$ is the characteristic wake width. Therefore, for a semi-infinite wind farm, we obtain
\begin{equation}\label{eq:u_dji_define}
\bar{u}_{d,m}(x_n)=\lim_{K\to\infty}C_{mn}\sum_{k=-K}^{K}\textrm{exp}\left(-\frac{\left(y_n-y_m-ks\right)^2}{2\sigma_{mn}^2}\right),
\end{equation}
where $C_{mn}$ is the maximum velocity deficit caused by each WT$_m$ at $x=x_n$, $k$ denotes the column number which varies from $-\infty$ to $\infty$, and $\sigma_{mn}$ is the width of WT$_m$'s wakes at $x=x_n$. Solving \eqref{eq:u_dji_define} gives
\begin{equation}\label{eq:app:u_dji}
    \bar{u}_{d,m}(x_n)=\sqrt{2\pi}C_{mn}\frac{\sigma_{mn}}{s_y}\vartheta_{3}\left[{\frac{\pi(y_n-y_m)}{s_y}},{\textrm{exp}\left(-2\pi^2\frac{\sigma_{mn}^2}{s_y^2}\right)}\right],
\end{equation}
where $\vartheta_{3}[z,q]$ is the Jacobi theta function defined as $1+2\sum_{l=1}^{\infty}q^{l^2}\cos(2lz)$ \citep{whittaker2020}. High-level programming languages (e.g., Python) may have a built-in function to compute the Jacobi theta function \citep{mpmath}. The series describing the Jacobi theta function  converges rather quickly, so as an approximation, one can alternatively compute the summation of the first few terms in the series. It is also noteworthy that $\vartheta_{3}[z\pm\pi,q]=\vartheta_{3}[z,q]$, so $y_n$ and $y_m$ in \eqref{eq:app:u_dji} can be the spanwise location of any turbines in WT$_n$ and WT$_m$, respectively.

By definition $U_{d,m}(x)=\langle\bar{u}\rangle_{d,m}(x)$, so we can write
\begin{equation}\label{eq:app:bar_u_dji_avg}
    U_{d,m}(x_n)=\lim_{K\to\infty}\frac{C_{mn}}{Ms_y}\sum_{k=-K}^{K}\int_{-\infty}^{+\infty}\textrm{exp}\left(-\frac{(y-y_m-ks_y)^2}{2\sigma^2_{mn}}\right)\textrm{d}y=\sqrt{2\pi}C_{mn}\frac{\sigma_{mn}}{s_y}.
\end{equation}
Therefore, from \eqref{eq:app:u_dji} and \eqref{eq:app:bar_u_dji_avg}, we conclude that
\begin{equation}\label{eq:app:u_dji2}
    \bar{u}_{d,m}(x_n)=U_{d,m}(x_n)\vartheta_{3}\left[{\frac{\pi(y_n-y_m)}{s_y}},{\textrm{exp}\left(-2\pi^2\frac{\sigma_{mn}^2}{s_y^2}\right)}\right].
\end{equation}
The wake width $\sigma_{mn}$ can be simplified by $k^{*}(x_n-x_m)+\sigma_0$, where the wake expansion $k^{*}$ is typically around 0.02-0.04 for offshore conditions and $\sigma_0$ is around 0.2-0.3 \citep{bastankhah14}. For simplicity, we assume $k^{*}=0.025$ and $\sigma_0=0.25$ to reduce  \eqref{eq:app:u_dji2} to
\begin{equation}\label{eq:app:u_dji3}
    \bar{u}_{d,m}(x_n)\approx U_{d,m}(x_n)\vartheta_{3}\left[{\frac{\pi(y_n-y_m)}{s_y}},{\textrm{exp}\left(-\frac{(x_n-x_m+10)^2}{80s_y^2}\right)}\right],
\end{equation}
Therefore from \eqref{eq:app:hub_height_velocity} and \eqref{eq:app:u_dji3}, we conclude
\begin{figure}
    \centering
    \begin{overpic}[width=.65\textwidth]{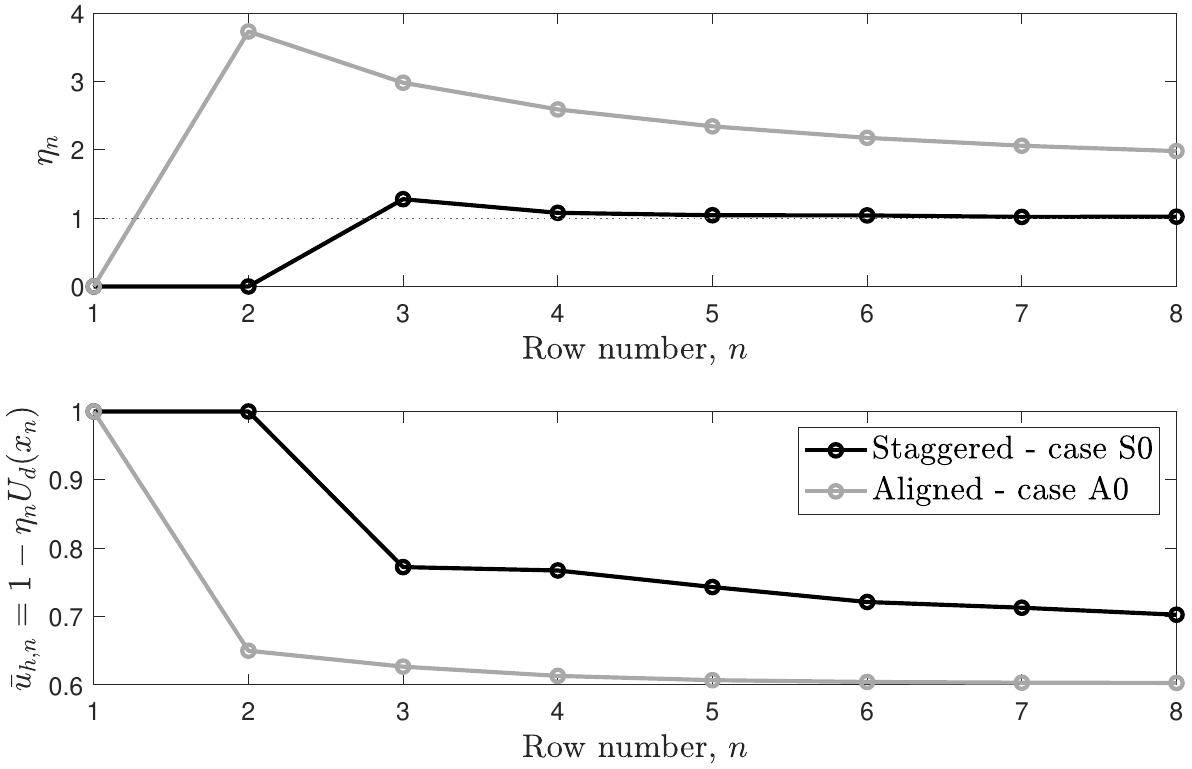}
       \put(-10,65){(a)} 
  \put(-10,30){(b)}
    \end{overpic}
    \caption{Variation of (a) local deficit coefficient  $\eta_n$ and (b) local hub-height velocity $\bar{u}_{h,n}$ with the row number $n$ for the Staggered Baseline (S0) and the Aligned baseline (A0) cases based on \eqref{eq:alpha}. }
    \label{fig:hub-height-velocity}
\end{figure}
\begin{equation}\label{eq:alpha}
  \eta_{n} =\frac{\sum\limits_{m=1}^{n-1} U_{d,m}(x_n)\vartheta_{3}\left[{\frac{\pi(y_n-y_m)}{s_y}},{\textrm{exp}\left(-\frac{(x_n-x_m+10)^2}{80s_y^2}\right)}\right]}
  {\sum\limits_{m=1}^{n-1}U_{d,m}(x_n)},      
\end{equation}     
where $n>1$ (for $n=1$, $\eta$ is zero), and $U_{d,m}(x)$ is defined in \eqref{eqs:final_form_Udn}. Figure \ref{fig:hub-height-velocity} shows the variation of the local deficit coefficient $\eta_n$ with the row number $n$ for both the Aligned Baseline (A0) and the Staggered Baseline (B0) wind farms. It is interesting to note that the value of $\eta_n$ is always greater than one for an aligned wind farm. This is expected as in this case turbines operate in full-waked conditions. Therefore, the local velocity deficit experienced by turbines is expected to always be larger than the laterally-averaged velocity deficit. The maximum value of $\eta$ occurs at the second row where there is the maximum level of heterogeneity in the farm. Further downstream, due to the wake expansion and flow mixing, the value of $\eta_n$ decays and approaches a constant value around two for this particular wind farm layout. For the staggered wind farm, however, we observe a very different behaviour. For the second row, the value of $\eta_n$ is almost zero. This is due to the fact that, WT$_2$s are not affected by the wake of WT$_1$s due to the staggered layout of the wind farm. Wake interactions only occur from the third row where there is a sudden increase in the value of $\eta_n$. The value of $\eta_n$ for the staggered wind farm approaches one for WT$_4$ and downwind rows. This suggests a fairly uniform distribution of streamwise velocity deep inside a staggered wind farm. In figure \ref{fig:hub-height-velocity}b, we can see how different distributions of the local velocity coefficient $\eta_n$ lead to different distribution of local hub-height velocity $\bar{u}_{h,n}$. The local hub height velocity is clearly higher for the staggered wind farm layout which leads to more power generation. The trend shown in figure \ref{fig:hub-height-velocity}b is similar to the data reported in other works \citep[e.g.,][]{chamorro-et-all2011,stieren2022impact}. We have only discussed aligned and staggered layouts here, but an important feature of our developed model is that it is generalisable, through variable local deficit coefficient $\eta_n$ developed in \eqref{eq:alpha}, to any conceivable wind farm layout provided the layout fulfills the requirements defined in section \ref{sec:wind_farm_definition}.

\subsection{Estimation of shear term $C_x$ and veer term $C_y$}\label{sec:Cx and Cy}
As discussed earlier in \eqref{eqs:velocity_defict}, $C_{x}=-\nu_{t,f} \partial^2 U_0/\partial z^2$ and $C_{y}=-\nu_{t,f} \partial^2 V_0/\partial z^2$. Based on \eqref{eqn:DANS_ekman},  $C_{x}$ and $C_{y}$ can be written as
\begin{align}
\begin{split}\label{Eqs:C_xC_y}
C_{x}&=-\frac{\nu_{t,f}}{\nu_{t,0}}f_c(V_g-V_0),\\
C_{y}&=\frac{\nu_{t,f}}{\nu_{t,0}}f_c(U_g-U_0).
\end{split}
\end{align}
As a first-order approximation, the vertical changes in the streamwise and spanwise velocities across the ABL are proportional to $G\cos\theta_0$ and $G\sin\theta_0$, respectively. Here, $G=\sqrt{U_g^2+V_g^2}$ is the geostrophic wind speed, and the cross-isobar angle $\theta_0$ is the angle between the wind direction on the ground surface and the geostrophic wind direction. One may thus write
\begin{align}
\begin{split}\label{Eqs:C_xC_y2}
U_g-U_0&=c_3G\cos\theta_0,\\
V_g-V_0&=c_3 G\sin\theta_0,
\end{split}
\end{align}
where $c_3$ is a constant. Values of $G$ and $\theta_0$ can be obtained from the widely-used geostrophic drag law (GDL), which relates surface properties (e.g., $z_0$ and $u_*$) to geostrophic wind speed on top of the ABL \citep[e.g.,][]{blackadar1968asymptotic,hess2002evaluating,van2020rossby}. For a neutrally-stratified ABL, the geostrophic drag law reads as \citep{liu2021geostrophic}
\begin{align}
\begin{split}\label{eq:geostrophic}
        A&=\ln\left(\frac{u_*}{z_{0,0}|f_c|}\right)-\frac{\kappa G}{u_*}\cos\theta_0,\\
B&=\mp \frac{\kappa G}{u_*}\sin\theta_0,
\end{split}
\end{align}
where $A$ and $B$ are universal empirical constants, and values of $A = 1.8$ and $B= 4.5$ used in the Wind Atlas Analysis and Application Program (WAsP) \citep{floors2018WASP} are adopted here. The minus sign on the right-hand side of the second equality in \eqref{eq:geostrophic} relates to the northern hemisphere where $f_c$ is positive, whereas the positive sign is for the southern hemisphere where $f_c$ is negative \citep{liu2021geostrophic}. From \eqref{Eqs:C_xC_y}, \eqref{Eqs:C_xC_y2} and \eqref{eq:geostrophic}, we can therefore approximate the values of $C_x$ and $C_y$ as follows:
\begin{align}
\begin{split}\label{eq:cx-cy-final}
C_{x}&=c_3|f_c|\frac{\nu_{t,f}}{\nu_{t,0}}\frac{u_*}{\kappa}B,\\
C_{y}&=c_3f_c\frac{\nu_{t,f}}{\nu_{t,0}}\frac{u_*}{\kappa}\left[\ln\left(\frac{u_*}{z_{0,0}|f_c|}\right)-A\right].
\end{split}
\end{align}
While for simplicity the conventional GDL relationship is used here to estimate $C_x$ and $C_y$, recent works \citep[e.g.,][]{liu2022ekman,narasimhan2023ekman} that describe the structure of Ekman boundary layer flows can be implemented in future works.
\section{Results and discussions}\label{sec:results}
\begin{table}
    \centering
    \begin{tabular}{m{5em} m{5em} m{5em}}
                $c_1$ & $c_2$ & $c_3$ \\ \hline 0.85 & 0.05 & 0.06
    \end{tabular}
    \caption{Empirical model coefficients used in this study.}
    \label{tab:model_coeff}
\end{table}
\begin{figure}
    \centering
    \includegraphics[width=.65\textwidth]{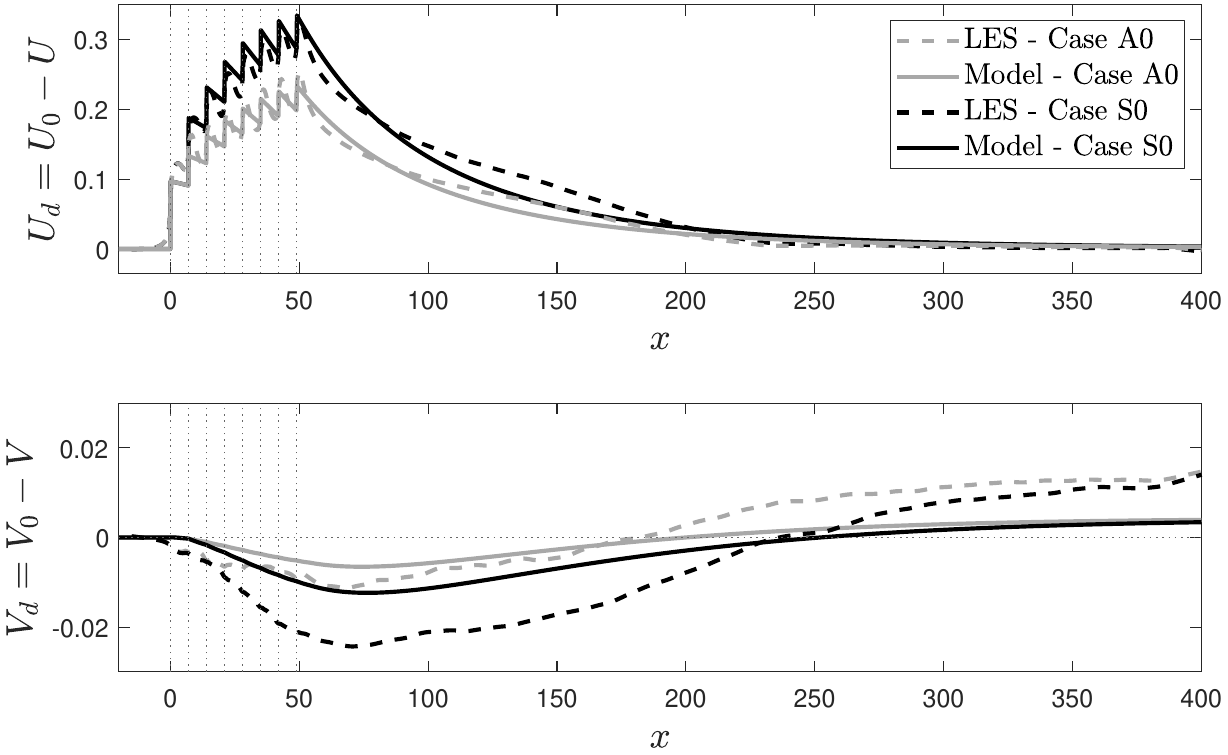}
    \caption{Variation of laterally-averaged streamwise velocity deficit $U_d$ and spanwise velocity deficit $V_d$ at the turbine hub height for the Aligned Baseline (A0) and the Staggered Baseline (S0) case.  The dashed lines show the LES data, and the solid lines show predictions of the developed model. Vertical dotted lines show the locations of wind turbine rows.}
    \label{fig:staggered}
\end{figure}

In this section, we compare predictions of the model developed in section \ref{sec:model_development} with the LES data elaborated in section \ref{sec:LES_setup}. The values of model coefficients, namely $c_1$, $c_2$ and $c_3$, used in this study are listed in table \ref{tab:model_coeff}. It is worth reminding that $c_1$ is the coefficient that is multiplied to $\nu_t=\nu_{t,0}+\nu_{t,f}$ in the final solution, while $c_2$ is the coefficient within $\nu_{t,f}$. An increase of either $c_1$ or $c_2$ increases the wake recovery rate, but $c_2$ is the one that quantifies the importance of $\nu_{t,f}$ over $\nu_{t,0}$. On the other hand, $c_3$ is the coefficient of $C_x$ (i.e., shear term in the streamwise momentum equation) and $C_y$ (i.e., veer term in the spanwise momentum equation). Given that $C_x$ is fairly small compared to other terms in the streamwise momentum equation, the main impact of $c_3$ is on predictions of the spanwise velocity deficit. The reported coefficients were found manually based on comparing model outputs with the LES data, but a more systematic approach based on an optimisation algorithm might provide more suitable coefficients. We must also note that although the coefficients suggested in table \ref{tab:model_coeff} provide satisfactory predictions for several cases studied here, future research is indeed required to examine whether these values are universal.  

First, we discuss the laterally-averaged velocity deficit for the Aligned Baseline (A0) case shown in figure \ref{fig:staggered}. For the streamwise direction, the figure shows a sudden jump in velocity deficit at each row, which is due to the turbine thrust force (i.e., $\langle\bar{f}_x\rangle$ in \eqref{eqs:velocity_defict}). The model underpredicts the velocity-deficit increase for the first row. This is likely due to the lack of modelling farm-scale blockage effects. The budget analysis in section \ref{sec:budget_analysis} showed a considerable flow deceleration in the vicinity of WT$_1$s due to farm-scale blockage effects. 
Both LES data and model predictions suggest that the velocity deficit jump due to turbine thrust forcing is significantly reduced after the first row. This is mainly due to the fact that as shown in \eqref{eqs:turbine_forcing}, the thrust force is proportional to the square of the local hub-height velocity, which is clearly lower for subsequent turbine rows. 

\begin{figure}
    \centering
    \includegraphics[width=.8\textwidth]{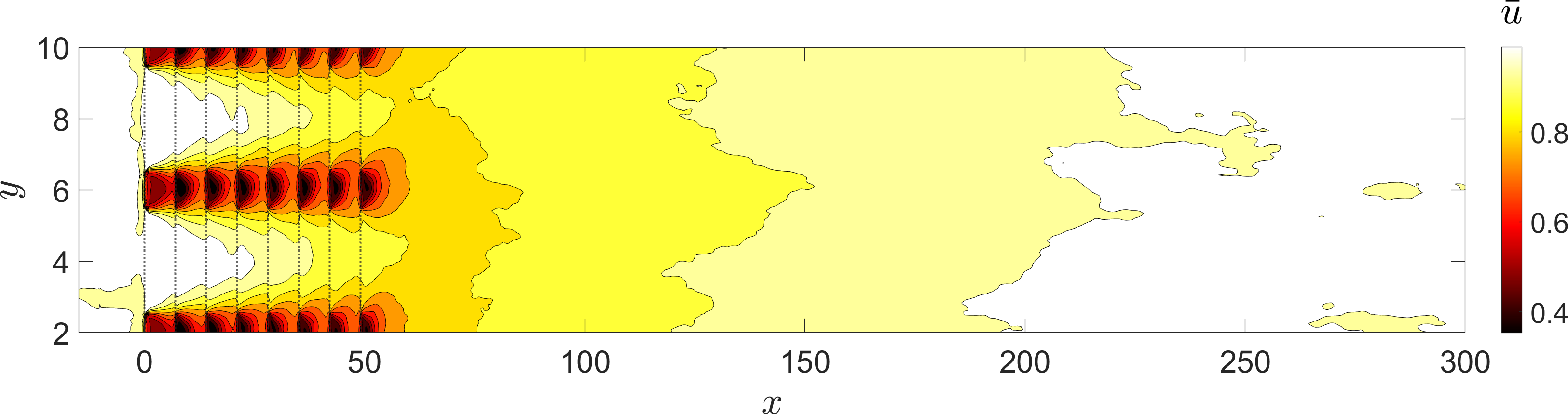}
    \caption{Contours of time-averaged streamwise velocity $\bar{u}$ on a horizontal plane at the turbine hub height for Aligned Baseline (A0) case. Vertical dotted lines denote the location of wind turbine rows.}
    \label{fig:turbine_farm_transition}
\end{figure}

After the last row of turbines, the velocity deficit diminishes rather rapidly in the farm near-wake region (e.g., for $x=50-100$). The primary reason for this fast recovery in the farm near-wake region is the large value of Reynolds stress gradient ($\partial \langle\overline{u'w'} \rangle\partial z$) as shown in figure  \ref{fig:budget-streamwise}. The dispersive stress gradient (i.e., ${\partial\langle\overline{u}''\overline{u}''\rangle}/{\partial x}$) is large immediately behind the wind farm, but it decays rapidly. As discussed previously, the dispersive stress quantifies the level of inhomogeneity in the flow. Right behind the wind farm, turbine wakes are not completely merged yet. This creates large velocity gradients over small length scales which in turn lead to higher turbulence generation. Further downstream, the turbine wakes merge and form a single farm wake. This is where the dispersive stress becomes negligible, and the gradient of Reynolds shear stress becomes the sole mechanism for the wake recovery. Merging of individual turbine wakes to form a single farm wake is evident in figure \ref{fig:turbine_farm_transition} that shows contours of the time-averaged streamwise velocity $\bar{u}$ at the hub height. This is conceptually similar to the transition of a wake array to a single wake that occurs downwind of multi-rotor turbines discussed in \cite{bastankhah2019multirotor}.

The developed model can capture the fast wake recovery in the farm near wake. In the momentum equation \eqref{eqs:velocity_defict}, the streamwise velocity-deficit reduction is mainly caused by the recovery term $c_1\nu_tU_d$. In the farm near wake, the streamwise velocity deficit $U_d$ is still fairly large. Moreover, the farm turbulent viscosity $\nu_{t,f}$ has its maximum value immediately after the wind farm as shown in figure \ref{fig:turbulent_viscosity}. Both of these contribute to a fast wake recovery in the farm near-wake region. In the farm far-wake region (e.g., for $x>150$), the velocity deficit clearly decays at a slower rate. According to the budget analysis in section \ref{sec:budget_analysis}, in this region, the advection term is in balance with the Reynolds stress gradient, whose effect is modelled by the recovery term $c_1\nu_tU_d$ in \eqref{eqs:velocity_defict}. However, both $\nu_t$ and $U_d$ are much smaller in this region which leads to a slower wake recovery rate. This explains the persistence of the wake over a very long distance. Figure \ref{fig:staggered} shows that the velocity deficit is still not negligible even after  twenty kilometers downstream of the wind farm (i.e., $x\approx 208$). It is also worth noting that for the LES cases studied here, the Coriolis term $f_cV_d$ is an order of magnitude smaller than other terms in the streamwise momentum equation \eqref{eqs:velocity_defict}. Moreover, while the shear term $C_x$ is not negligible, it is much smaller than the recovery term $c_1\nu_tU_d$ for this configuration. However, this is not case for the spanwise momentum equation as elaborated in the following.   
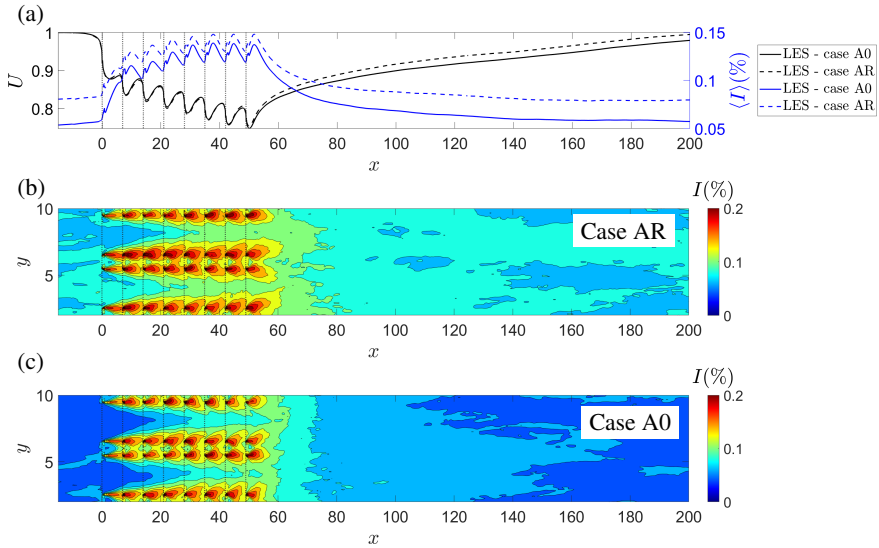
\begin{figure}
 \centerline{\begin{overpic}[width=.85\textwidth]{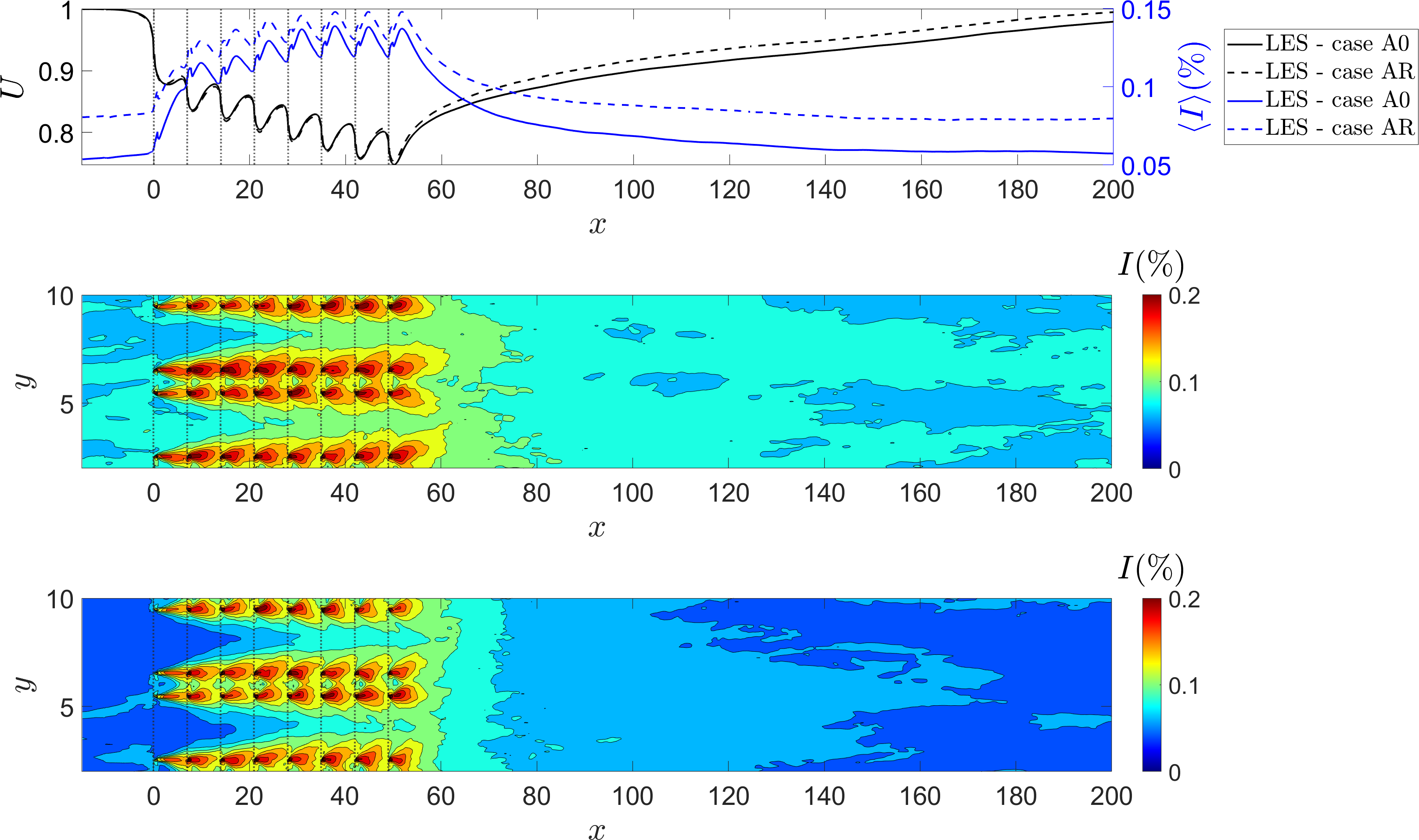}
 \put(1,60){(a)} 
  \put(1,40){(b)}
  \put(65,35){\colorbox{White}{Case AR}}
   \put(1,20){(c)}
   \put(66,13){\colorbox{White}{Case A0}}
\end{overpic}}
\caption{(a) Variation of laterally-averaged streamwise turbulence intensity $\langle I \rangle$ and streamwise velocity $U$ with downwind distance $x$ for both cases of Aligned Baseline (A0) and Aligned Rough (AR). Contours of laterally-averaged streamwise turbulence intensity $\langle I \rangle$ on a horizontal plane at the hub height for (b) the AR case and (c) for the A0 case.}
\label{fig:roughness_turbulence}
\end{figure}

The spanwise turbine forcing term $\langle \bar{f}_y\rangle$ defined in \eqref{eqs:turbine_forcing} is zero for the LES data given that the yaw angle $\gamma$ is zero. The Coriolis term in the spanwise momentum equation \eqref{eqs:velocity_defict} however is important because it is proportional to $U_d$, which has a considerable value especially within the farm. Therefore, the Coriolis term increases in the farm with an increase of streamwise velocity deficit. According to \eqref{eqs:velocity_defict}, the Coriolis term with $U_d>0$ leads to a negative $V_d$ (i.e., positive $V$) in the northern hemisphere, where $f_c>0$. This is described in the literature \citep[e.g.,][]{vanderLaan2017WhyHemisphere} as an anticlockwise deflection based on a view from top. The initial anticlockwise deflection of the wake is shown in figure \ref{fig:staggered}. Apart from the streamwise wake deficit, the farm-induced turbulence also impacts the distribution of spanwise velocity deficit. In particular, the increase of farm turbulent viscosity $\nu_{t,f}$ has two important effects. It increases both the recovery term $c_1\nu_t V_d$ and the veer term $C_y$ in \eqref{eqs:velocity_defict}. Both of these effects reduce the initial anticlockwise deflection of the wake such that after some downwind distances, the effect of the veer term $C_y$ becomes dominant, and the direction of the wake deflection changes to clockwise (i.e., $V_d>0$). The change in the direction of wake deflection was also reported in \cite{gadde2019Stevens}. Ultimately, very far downstream (not shown here), $C_y$ goes to zero as $\nu_{t,f}$ goes to zero, and therefore the spanwise velocity deficit is completely diminished by the recovery term. This conflicting behaviours of Coriolis and veer on the spanwise velocity coexist as the latter is the consequence of the former. However, depending on atmospheric conditions and simulation settings, their relative magnitude with respect to each other could be different at different downstream positions. For instance, the succeeding clockwise deflection observed in figure \ref{fig:staggered} can completely mask the initial anticlockwise deflection. This may happen either in the case of a strong wind veer which typically occurs in thermally-stable boundary layers, or if the wind farm generates a high amount of turbulence. On the other hand, an anticlockwise deflection is mainly observed if both incoming wind veer and farm-generated turbulence are relatively small, or if only the farm near wake is of interest. This may explain why some prior works observed a clockwise deflection \citep[e.g.,][]{abkar2016wake,vanderLaan2017WhyHemisphere,eriksson2019ivanell,nouri2020coriolis}, whereas others reported an anticlockwise deflection \citep[e.g.,][]{dorenkamper2015impact,allaerts2017,frank2023coriolis}.

\begin{figure}
    \centering
    \includegraphics[width=.65\textwidth]{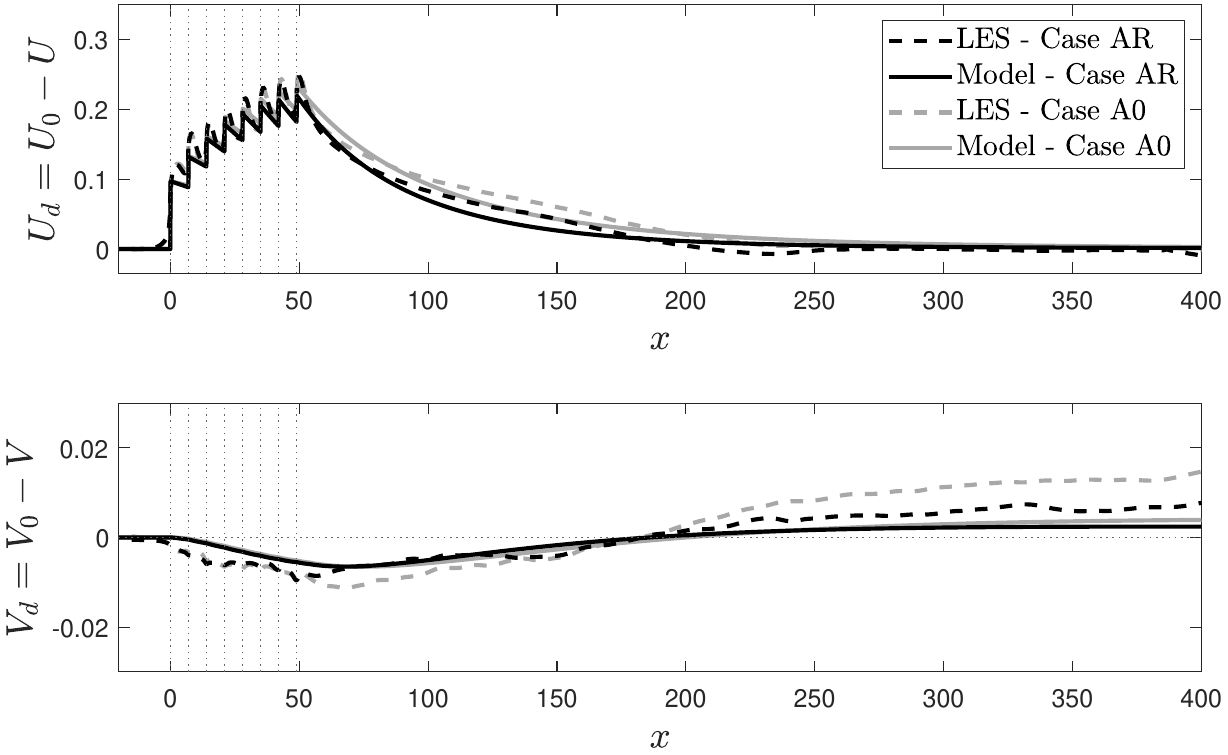}
    \caption{
Variation of the laterally-averaged streamwise velocity deficit $U_d$ and the spanwise velocity deficit $V_d$ at the turbine hub height for the Aligned Rough (AR) case. For comparison, the data for the Aligned Baseline (A0) case are also shown. The dashed lines show the LES data, and the solid lines show predictions of the developed model. Vertical dotted lines show the locations of wind turbine rows.}
    \label{fig:roughness_model}
\end{figure}

Overall, the agreement of the model predictions with the LES data is satisfactory for the streamwise velocity deficit $U_d$. The developed model can also successfully capture the overall trend for the spanwise velocity deficit $V_d$. In agreement with the LES data, it predicts the initial anticlockwise deflection followed by a clockwise deflection. However, we should note that figure \ref{fig:staggered} shows some differences in the magnitude of $V_d$ especially in the far wake. The value of $V_d$ is one order of magnitude smaller than $U_d$ for these LES data. While predicting $V_d$ with such small values in these cases can be challenging, the model's prediction for $V_d$ compared to the LES data remains within $1\%$ of $\mathcal{U}_h$, and the difference in wind direction $\varphi$ prediction is also within 1 degree (not shown here). In this work, we only examined neutral boundary layers where wind veer is not strong. For thermally-stable boundary layers, the cross-isobar angle $\theta_0$ in \eqref{Eqs:C_xC_y2} is expected to be considerably larger \citep{pena2014veer}, which in turn increases the value of $C_y$. It is therefore of great interest to extend the model to thermally-stratified boundary layers in future works, where the deflection of the wake due to the wind veer is expected to be significantly larger. The discrepancy in $V_d$ observed between the LES data and the model can be also partly explained by the linearisation of momentum equations. In order to mathematically solve the equations, we replaced $U$ with $U_0=1$ in the advection term in \eqref{eqn:DANS_simplified}. In the regions where the difference between $U$ and $U_0$ is not negligible (i.e., within the wind farm), this assumption leads to an underestimation of the velocity deficit. The error introduced by this assumption is expected to be more evident in spanwise velocity predictions given their small values.

Next, we discuss the effect of wind farm layout on the farm flow distribution. The variation of the velocity deficit for the Staggered Baseline (S0) case is shown in figure \ref{fig:staggered}. 
 The main notable difference in the streamwise velocity deficit between the two wind farm layouts (aligned vs staggered) is the fact that, from the second row of turbines, the jump in $U_d$ due to turbine forcing is clearly larger for the staggered wind farm, in agreement with prior studies \citep[e.g.,][]{stevens2016effects,stieren2022impact}. This leads to the maximum velocity deficit $U_d$ of 0.32 for the staggered wind farm, which is 28\% larger than the one for the aligned wind farm. 
As discussed in section \ref{sec:local_hub_height_ve}, turbines within the staggered wind farm experience a larger local velocity which according to \eqref{eqs:turbine_forcing} leads to a larger value of turbine thrust force. This highlights the importance of the local deficit coefficient $\eta$ discussed in section \ref{sec:local_hub_height_ve}
 and implemented in the model \eqref{eqs:final_form}. Without this coefficient, model predictions will be the same for both cases of $A0$ and $S0$, which is clearly unrealistic according to the LES data. The enhanced turbulence mixing caused by large values of velocity deficit in the staggered wind farm accelerates the wake recovery downstream. At about $x=200$, the streamwise velocity deficit becomes approximately equal in the wake of both wind farms. The faster recovery of the wake for the staggered wind farm is captured in the model given that the recovery term in \eqref{eqs:velocity_defict} depends on the value of $U_d$. 

The LES data also shows that the spanwise velocity deficit is larger for the staggered wind farm in comparison with the aligned wind farm. As discussed earlier, the Coriolis term in the spanwise momentum equation \eqref{eqs:velocity_defict} directly depends on $U_d$. Therefore, one expects to observe a larger value of $V_d$ for the staggered wind farm. Due to the initial large anticlockwise wake deflection, the transition to the clockwise deflection occurs at a later downstream position for the staggered wind farm.  However, the wake deflection for both wind farms eventually approaches the same value in the very far wake region. Consistent with the LES data, the model predicts a larger negative peak of the spanwise velocity deficit in this case. The agreement is however less satisfactory for the staggered layout. The larger streamwise velocity deficit for the S0 case may exacerbate the error introduced by the linearisation of the momentum equations discussed earlier. Nonetheless, the results are still within a difference of $1\%\mathcal{U}_h$.  

\begin{figure}
    \centering
    \includegraphics[width=.65\textwidth]{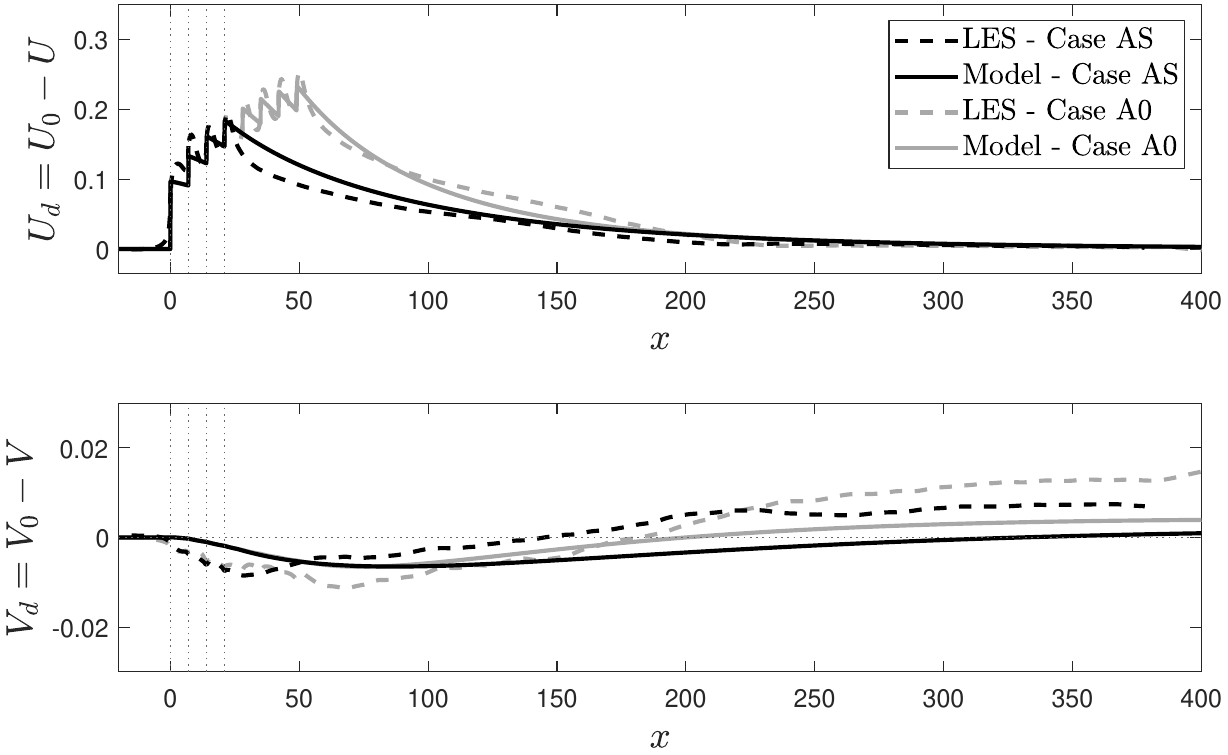}
    \caption{
Variation of laterally-averaged streamwise velocity deficit $U_d$ and spanwise velocity deficit $V_d$ at the turbine hub height for the Aligned Short (AS) case. For comparison, the data for the Aligned Baseline (A0) case are also shown. The dashed lines show the LES data, and the solid lines show predictions of the developed model. Vertical dotted lines show the locations of wind turbine rows for the AS case.}
    \label{fig:length_farm}
\end{figure}

Next, we study the effect of surface roughness on the evolution of wind farm wakes. We compare the results for the two cases of the Aligned Baseline (Case A0) with $z_0=2\times 10^{-4}\textrm{m}/D$ and the Aligned Rough (Case AR) with $z_0=2\times 10^{-2}\textrm{m}/D$. Figure \ref{fig:roughness_turbulence} compares the LES data for these two cases. Figure \ref{fig:roughness_turbulence}(a) shows that, as expected, the incoming streamwise turbulence intensity $\langle I\rangle$ is clearly larger for the case with the higher roughness. However, the difference in the turbulence level between the two cases become less clear within the wind farm, especially towards the end of the wind farm as shown in figure \ref{fig:roughness_turbulence}(a-c). In other words, these data suggest that the turbulence added by the wind farm is negatively proportional to the ambient turbulence level. This is consistent with the empirical relation of \cite{crespo1996} for the added wake turbulence, as highlighted in \cite{zehtabiyan2023short}. In addition, it is important to note that the higher turbulence level in the AR case promotes the wake recovery after each turbine row which in turn increases the incoming wind speed for the next turbine. This however increases the velocity deficit jump that occurs at the next turbine row according to \eqref{eqs:turbine_forcing}. This is why despite the difference in the incoming turbulence level, the two cases have fairly similar velocity distribution within the wind farm as shown in \ref{fig:roughness_turbulence}(a). This suggests that the wind farm region is highly dominated by the turbulence generated by the wind farm, and it seems to be less dependant on the incoming turbulence level. However, in the farm wake where the turbulence added by the farm gradually diminishes, the impact of the ambient turbulence becomes more important. Figure \ref{fig:roughness_turbulence}(a) shows that the wake of the AR case recovers faster than the one of the A0 case. Model predictions in comparison with the LES data for these two cases are depicted in figure \ref{fig:roughness_model}, which overall shows a good agreement for the streamwise velocity deficit. The model predicts the further reduction of the velocity deficit in the wake of the wind farm for the AR case. For the spanwise velocity deficit, results seem to be fairly similar for the two cases. The only notable difference can be observed in the far-wake region of the farm where the spanwise deficit for the AR case is less than the one for A0 case. Similar to the streamwise velocity deficit, the smaller spanwise deficit for the AR case is due to the larger value of ambient turbulence and thereby faster wake recovery.

\begin{figure}
    \centering
    \includegraphics[width=.65\textwidth]{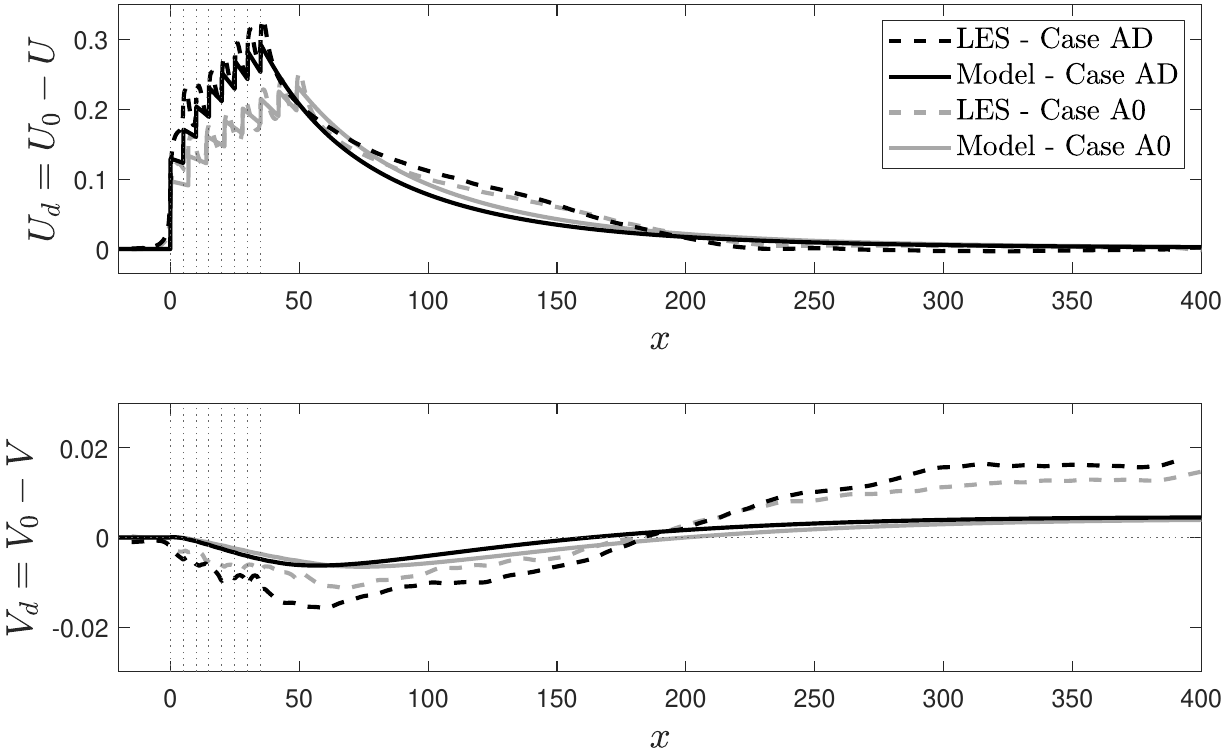}
    \caption{Variation of laterally-averaged streamwise velocity deficit $U_d$ and spanwise velocity deficit $V_d$ at the turbine hub height for the Aligned Dense (AD) case. For comparison, the data for the Aligned Baseline (A0) case are also shown. The dashed lines show the LES data, and the solid lines show predictions of the developed model. Vertical dotted lines show the locations of wind turbine rows for the AD case.}
    \label{fig:density_farm}
\end{figure}

\begin{figure}
    \centering
    \includegraphics[width=.65\textwidth]{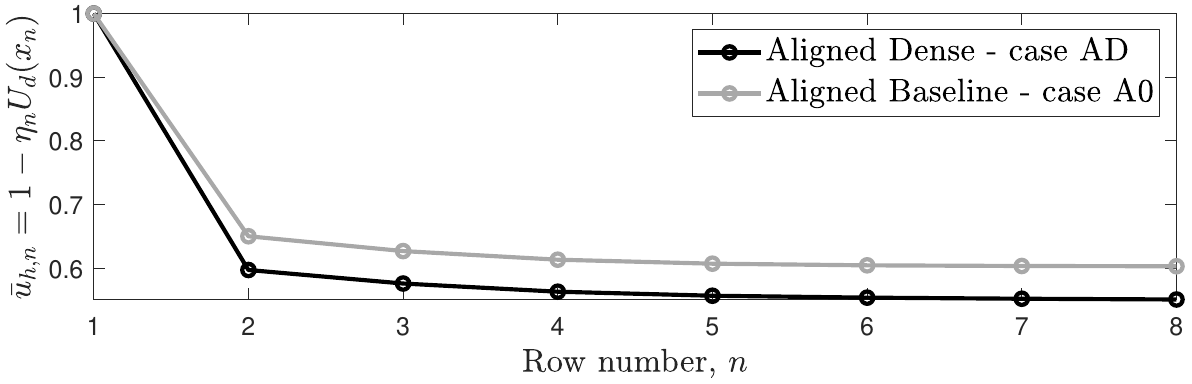}
    \caption{Variation of the local hub-height velocity $\bar{u}_{h,n}$ as a function of turbine row number $n$ for the Aligned Dense (AD) and the Aligned Baseline (A0) cases based on the proposed model.}
    \label{fig:hub_turbine-density}
\end{figure}

Figures \ref{fig:length_farm} and \ref{fig:density_farm} show respectively the effect of wind farm length and turbine spacing on the wake evolution for both the LES and the model. Figure \ref{fig:length_farm} shows that the streamwise velocity deficit is smaller for the short wind farm. This can be simply explained by the fact that there are fewer turbine rows in this wind farm and thereby less thrust forcing acting on the incoming wind. In figure \ref{fig:density_farm}, however, the number of turbine rows are the same, and they differ in both streamwise $s_x$ and spanwise $s_y$ inter-turbine spacing as shown in table \ref{tab:runs}. It is interesting to note that reducing $s_x$ and $s_y$ has multifaceted effects on the laterally-averaged streamwise velocity deficit. The reduction of $s_y$ directly increases the velocity deficit, because by reducing $s_y$, turbine wakes occupy a larger potion of the flow field. On the other hand, with a reduction of $s_x$, the local incoming velocity for downwind turbines decreases as shown in figure \ref{fig:hub_turbine-density}. This reduces the amount of turbine thrust force, which is expected to decrease the total velocity deficit generated by wind turbines. However, the turbine spacing also affects the rate of wake recovery. Decreasing both $s_x$ and $s_y$ increases the turbulent farm velocity scale $\hat{u}_f$, which leads to a larger $\nu_{t,f}$ and thus faster wake recovery. Therefore, knowledge on the relative importance of each of these factors is needed for each case in order to predict the overall impact of changing the turbine spacing on the wake velocity deficit. Figure \ref{fig:density_farm} shows that for this case the impact of $s_y$ change on the velocity deficit is initially dominant as the velocity deficit within the AD farm is much larger than the one of the A0 farm. Further downstream, however, wakes of both wind farms experience a fairly similar level of streamwise velocity deficit. The spanwise velocity deficit is also approximately similar in both cases.

\section{Summary}\label{sec:summary}
The aim of this work is to develop a new physics-based one-dimensional model to predict the variation of laterally-averaged streamwise and spanwise velocities in the wake of a wind farm at the turbine hub-height level. Through a budget analysis based on the LES data of semi-infinite wind farms, dominant terms in the momentum equations were identified. This led to an approximate form of the momentum equations where the sum of the Coriolis force, the divergence of the Reynolds stresses, and the turbine thrust force are in balance with the change in momentum by advection. 

The linearised versions of the approximate form of the momentum equations in both the streamwise and spanwise directions were then mathematically solved to obtain the the proposed model stated in \eqref{eqs:final_form}. To derive this solution, the turbulent viscosity hypothesis was used to model the Reynolds shear stresses. The turbulent viscosity $\nu_t$ was decomposed into the ambient turbulent viscosity $\nu_{t,0}$ and the farm turbulent viscosity $\nu_{t,f}(x)$, where the latter changes with $x$. The dependency of  $\nu_{t,f}$ to $x$ was modelled using a velocity scale proportional to the turbine forcing per unit area, and a length scale proportional to the thickness of the internal boundary layer $\delta$. The proposed model importantly accounts for the fact that wind farms with different layouts may generate noticeably different wakes. This is mainly due to the fact that the local incoming velocity experienced by wind turbines, and thus the thrust force that is exerted by the wind turbines on the airflow depends on the farm layout. A geometric parameter called the local deficit coefficient $\eta$ was introduced to relate the local velocity deficit at the rotor-centre of wind turbines to the laterally-averaged velocity deficit at the same streamwise position. Moreover, the gradients of the incoming wind shear and wind veer appeared in terms $C_x$ and $C_y$ of the final solution \eqref{eqs:final_form} and were estimated using the Geostrophic drag law. 

The model predictions are compared to LES data for five different cases to capture the response of the model to various changes in farm operating conditions. The Coriolis and shear terms in the governing equation \eqref{eqs:velocity_defict} for the streamwise direction are relatively small. Therefore, the change in the streamwise velocity is mainly determined by how different operating conditions influence the turbine forcing term and the wake recovery term in \eqref{eqs:velocity_defict}. In general, a higher local incoming velocity increases the turbine forcing term, which leads to a higher velocity deficit. The wake recovery rate on the other hand is increased with an increase of either the turbulent viscosity or the velocity deficit. Therefore, the overall impact of changing a parameter such as inter-turbine spacing or incoming turbulence is not trivial, and these changes may lead to counteracting effects on the streamwise velocity deficit. Overall, our results showed that the proposed model is able to acceptably predict the variation of streamwise velocity deficit for the different cases studied here. For the spanwise velocity deficit, turbine forces were not present in the LES data (i.e., $\bar{f}_y=0$ as $\gamma=0$ for all turbines). However, in addition to the recovery term, the two Coriolis and veer terms in \eqref{eqs:velocity_defict} are of great importance. Our results showed that the Coriolis term initially introduces an anticlockwise deflection (i.e., deflection to the left) in the Northern Hemisphere. Further downstream, the veer term becomes dominant and introduces a clockwise deflection (i.e., deflection to the right) in the Northern Hemisphere. The former is the direct effect of the Coriolis force, whereas the latter is present due to (i) an indirect effect of the Coriolis force through wind veer, and (ii) the farm-generated turbulence. The total value and direction of the wake deflection due to the Coriolis force therefore depends on the streamwise location and more importantly on the relative magnitude of these two counteracting terms with respect to each other. 

This work serves as the first study to model the wake of a semi-infinite wind farm. While only aligned and staggered layouts were modelled by our LESs, the developed model is capable of modelling different layouts. Therefore, future research can implement the model to study a wider range of layout configurations. Another interesting area of future research is to study the effect of yaw offset on the farm wake deflection. While this is not studied here, the effect of yaw angle is already incorporated in the model and can be used in future works. It is also of special interest to study how atmospheric thermal stratification may affect the turbulent viscosity and also our estimation for the shear and veer terms. Finally, it is important to remind ourselves that by definition the model assumes a wind farm that extends infinitely in the lateral direction. Therefore, model predictions only resemble the flow in the wake centre of a wide finite farm where side effects are deemed to be insignificant. More research is thus essential to quantify the impact of lateral flow entrainment and side effects. Moreover, the dimensional analysis used to simplify the vertical gradient of velocities in the recovery term of \eqref{eqs:velocity_defict} should be scrutinised in more detail. This simplification may imply that the cross-stream length-scale associated with the vertical flow shear remains comparable to the rotor diameter. This may not be necessarily true especially in the far wake of a wind farm.




\backsection[Funding]{This work was supported by Uppsala-Durham Seedcorn funding under the project entitled "systematic prediction of wind farm wakes: an emerging challenging in offshore wind sector". The LES simulations were run using resources provided by the Swedish National Infrastructure for Computing (SNIC).}

\backsection[Declaration of interests]{The authors report no conflict of interest.}




\bibliographystyle{jfm}
\bibliography{reference1,reference2}

\end{document}